\documentclass{aa}
\usepackage{aabib}
\usepackage[dvips]{graphics}
\usepackage{times}
\usepackage{amsfonts}
\usepackage{amsmath}
\usepackage{rotating}
\def\degr{$^{\circ}$}
\def\Msun{M$_{\odot}$}
\begin{document}
\thesaurus{03.13.2, 05.01.1, 08.02.4, 08.12.1, 08.06.3}
\title{Re-processing the Hipparcos Transit Data and Intermediate Astrometric Data
of spectroscopic binaries.\\
I. Ba, CH and Tc-poor S stars\thanks{Based on observations from the Hipparcos
astrometric satellite operated by the European Space Agency (ESA 1997)}}
\titlerunning{Hipparcos parameters of Ba, CH and S stars} 
\author{D.~Pourbaix\thanks{ESA-PRODEX contract 
13318/98/NL/VJ}\fnmsep\thanks{Scientific Collaborator, F.N.R.S., Belgium}\and 
A.~Jorissen\thanks{Research Associate, F.N.R.S., Belgium}} 
\institute{Institut d'Astronomie et d'Astrophysique, Universit\'e Libre de 
Bruxelles, C.P.~226, Boulevard du Triomphe, B-1050 Bruxelles, Belgium}
\date{Received date; accepted date} 
\offprints{pourbaix@astro.ulb.ac.be}
\maketitle
\begin{abstract}
Only 235 entries were processed as astrometric binaries with orbits in the
Hipparcos and Tycho Catalogue (\cite{Hipparcos}).  However, the
Intermediate Astrometric Data (IAD) and Transit Data (TD) made
available by ESA make it possible to re-process the stars that turned
out to be spectroscopic binaries after the completion of the
Catalogue.  This paper illustrates how TD and IAD may be used in
conjunction with the orbital parameters of spectroscopic binaries to
derive astrometric parameters.  The five astrometric and four orbital
parameters (not already known from the spectroscopic orbit) are
derived by minimizing an objective function ($\chi^2$) with an
algorithm of global optimization.  This code has been applied to 81
systems for which spectroscopic orbits became available recently and
that belong to various families of chemically-peculiar red giants
(namely, dwarf barium stars, strong and mild barium stars, CH stars,
and Tc-poor S stars).  Among these 81 systems, 23 yield reliable
astrometric orbits.  These 23 systems make it possible to evaluate
on real data the so-called `cosmic error' described by
\cite*{Wielen-1997}, namely the fact that an unrecognized orbital motion
introduces a systematic error on the proper motion.  Comparison of the
proper motion from the Hipparcos catalogue with that re-derived in the
present work indicates that the former are indeed far off the present
value for binaries with periods in the range 3 to $\sim$ 8 years.
Hipparcos parallaxes of unrecognized spectroscopic binaries 
turn out to be reliable, except for systems with
periods close to 1 year, as expected.  Finally, we show that, even when
a complete orbital revolution was observed by Hipparcos, the
inclination is unfortunately seldom precise.
\end{abstract}
\section{Introduction}
The chemical anomalies observed in several classes of late-type 
stars are due to mass transfer across a binary system. This 
scenario, first suggested by \cite*{McClure-1983:b}, holds for dwarf and 
giant barium stars (Pop.I G and K stars with overabundance of 
carbon and heavy elements like Sr and Ba produced by the s-process 
of nucleosynthesis), CH stars (the Pop.II analogs of Ba stars), and 
extrinsic S stars (late-type giants characterized by ZrO bands and 
no Tc lines, an element with no stable isotopes).  
The set of spectroscopic orbits available for those chemically-peculiar 
red stars (CPRS) has been considerably enlarged 
thanks to a decade-long radial-velocity monitoring. Such orbits are now 
available for 63 giant barium stars and 19 extrinsic 
S stars (\cite{Jorissen-1998}), 21 dwarf barium and subgiant CH stars 
(\cite{McClure-1997:a};North, priv. comm.) and 8 CH stars (\cite{McClure-1990}).
Most of them were not available to the Hipparcos reduction consortia.  The 
astrometric parameters (parallaxes and proper motions) published in the 
Hipparcos catalogue (\cite{Hipparcos}) were therefore derived from a single 
star solution (the so-called `5-parameter model'). Since many of these 
CPRS have orbital periods in the   
range 1 -- 5 yr, their unaccounted orbital motion may seriously 
confuse the parallax and proper motion as determined by the 
Hipparcos consortia.  Thus, unrecognized orbital motions might introduce
a so-called `cosmic error' (of a systematic nature) in the proper motion 
(\cite{Wielen-1997:b,Wielen-1999}).  Parallaxes may also possibly be in 
error, especially for binaries with orbital periods close to 1~yr. 

The conclusions of previous studies inferring the kinematics and absolute 
magnitudes of CPRS (\cite{Bergeat-1997,Mennessier-1997,VanEck-1998}) 
from the data of the Hipparcos catalogue may possibly be affected by 
these systematic errors. It is therefore of importance to re-evaluate the 
astrometric parameters of CPRS with a model correctly accounting for the 
orbital motion (the so-called `12-parameter model', reducing to 9
parameters if the spectroscopic orbital parameters $P, T$ and $e$ are
fixed), and to compare these values with those from the Hipparcos 
catalogue to identify the stars most affected by these systematic errors. 

This comparison should allow to assess the reliability of the 
kinematic properties and absolute magnitudes of CPRS formerly 
derived from the Hipparcos-catalogue values. As a by-product, the 
astrometric orbit has been obtained in many cases, leading to an 
estimate of the masses through the ratio $M_2/[1+(M_1/M_2)]^2\equiv f(M)
/\sin^3i$ (derived by combining the orbital inclination from the astrometric 
orbit with the mass function $f(M)$ from the spectroscopic orbit when 
only one spectrum is observable).  

From the operational point of view, the numerical re-processing of the raw 
Hipparcos data (Intermediate Astrometric Data or Transit Data; see 
Sect.~\ref{Sect:IAD}) presented in this paper is innovative on the following
points: (i) a global optimization technique is used to minimize the objective 
function in the 12-parameter (or, for spectroscopic binaries, 9-parameter) 
space, and (ii) a change of independent variable allows to derive always 
positive parallaxes, contrarily to the situation prevailing in the Hipparcos 
catalogue.  The tool developed here for CPRS may in the future be applied to 
any other binary star to re-process the Hipparcos data when new orbits appear.
The second paper of this series will be devoted to the re-processing of 
spectroscopic binaries involving a giant component (\cite{Boffin-1993}).  
Another is devoted to Procyon (\cite{Girard-2000}).  We encourage interested 
readers to communicate to the authors other spectroscopic binaries whose 
Hipparcos data would need to be reprocessed.

\section{Numerical methods}
\label{Sect:numerical}

In order to allow some re-processing of the Hipparcos data, the 
Hipparcos consortia provided the users with two kinds of data: the 
{\it Transit Data} (TD) and the {\it Intermediate Astrometric Data} 
(IAD).  TD (\cite{Quist-1999}) are a by-product of the analysis performed by 
the NDAC consortium, and merge astrometric and photometric data. They 
basically provide the signal as modulated by the grid in front of the 
detector. They are therefore especially well suited to the analysis of 
{\it visual} double or multiple systems, since the separation, position 
angle and magnitudes of the various components may in principle be 
extracted from the TD. For that reason, TD are not provided for all 
Hipparcos entries, but only for those stars known or suspected of 
being double or multiple systems.  IAD are lower level data which provide 
astrometric information only (the star's abscissa along a reference 
great circle, and the pole of that circle on the sky; 
\cite{vanLeeuwen-1998}), but for every Hipparcos entry.  
In the following, we describe how IAD and TD (when available) have 
been used to derive the astrometric parameters of the CPRS that have 
a known spectroscopic orbit. When TD are available, the astrometric 
parameters may be derived independently from the two data sets, and 
allow an interesting internal consistency check. 
In both cases, the method proceeds along the following steps: 
\begin{itemize}
\item definition of an objective function to minimize
\item computation of the objective function from the 
astrometric parameters
\item global minimization of the objective function with the {\em simulated 
annealing} algorithm
\item local minimization of the objective function (with a quasi-Newton
method)
\item comparison of the values of the objective function for the 
Hipparcos 5-parameter model  and for our 9-parameter model, 
and application of an F-test to evaluate whether the 9-parameter
model is significantly better than the Hipparcos solution
\item check of the astrophysical consistency  of the 9-parameter 
solution from some {\it a priori} knowledge of the properties of the 
system (i.e., component masses)
\end{itemize}
These steps are discussed in turn in the remaining of this section, 
first for IAD, and then for TD.

\subsection{Intermediate Astrometric Data} 
\label{Sect:IAD}
\subsubsection{Objective function}
\label{Sect:IAD:OF}
The first step in any data-fitting problem is to set up an objective 
function in order to compare different solutions (corresponding 
to different values of the model parameters). This function is 
usually constructed in such a way that its lowest value corresponds 
to the best solution. The problem then reduces to minimizing that function.
The IAD already include some corrections (e.g. aberration and 
satellite-attitude corrections) and they are no longer {\em stricto sensu}
raw data.  They provide the abscissa residuals along a reference great 
circle for the Hipparcos 5-parameter solution (i.e. $\alpha_0, \delta_0, 
\varpi, \mu_{\alpha}, \mu_{\delta}$, respectively: the position in right 
ascension and declination in the equinox 2000.0 at the epoch 1991.25, the 
parallax, the proper motions in right ascension and declination; following 
the practice of the Hipparcos catalogue, $\alpha_0^* = \alpha_0 \cos\delta$ 
and $\mu_{\alpha}^*= \mu_{\alpha} \cos\delta$ will be used instead of 
$\alpha_0$ and $\mu_{\alpha}$).
By slightly modifying these 5 parameters and by adding 7 orbital parameters 
(\cite{PrDoSt}), the aim is to further
reduce the abscissa residuals $\Delta v$ below the values obtained from the
Hipparcos 5-parameter model. 
In the case of the Hipparcos Intermediate Astrometric Data, the 
objective function is the $\chi^2$ expressed by Eq.~17.12 of Vol.3
in \cite*{Hipparcos}:
 \begin{equation}
\chi^2=(\Delta v-\sum_{k=1}^M\frac{\partial v}{\partial p_k}\Delta 
p_k)^{\mbox{t}}V^{-1}(\Delta v-\sum_{k=1}^M\frac{\partial 
v}{\partial p_k}\Delta p_k), \label{chi2}
\end{equation}
where the superscript $t$ means transposed, $\Delta v$ are the 
abscissa residuals provided by the IAD file and corresponding to 
the Hipparcos 5-parameter solution, and $\frac{\partial v}{\partial 
p_k}$ is the partial derivative of the abscissa with respect to the 
$k$-th parameter expressing how the abscissa residual varies 
when a correction $\Delta p_k$ is applied to the value of the $k$-th 
parameter with respect to the Hipparcos solution. 
$M$ is the number of parameters retained in the solution, and $V^{-
1}$ is the inverse of the 
covariance matrix of the observations 
given by: \[
V=\left(
\begin{array}{cccc}
V_1 & 0 & \cdots & 0\\
0 & V_2 & \ddots & \vdots\\
\vdots & \ddots & \ddots & 0 \\
0 & \cdots & 0 & V_n
\end{array}
\right)
\]
with
\begin{equation}
\label{Eq:sigma}
V_j=\left(
\begin{array}{cc}
\sigma_{F_j}^2 & \rho\sigma_{F_j}\sigma_{N_j} \\ 
\rho\sigma_{F_j}\sigma_{N_j} & \sigma_{N_j}^2 \end{array}
\right)
\end{equation}
If an observation $j$ was processed by the two consortia (FAST and 
NDAC), the residuals obtained by the two reduction consortia are 
correlated and $V_j$ is the $2\times 2$ variance-covariance 
matrix for measurement $j$.   On the other hand, when 
only one consortium processed observation $j$, $V_j$ reduces to one 
number\footnote{In order to have a unique expression for $\chi^2$ 
regardless of the dataset available (FAST, NDAC or both), we always 
write $V_j$ as a 2 by 2 matrix.  When, for whatever reason,  a measurement
from only one consortium  is used, that observation is duplicated and its 
weight divided by two ($V_j=\mbox{diag}(2\sigma_j^2,2\sigma_j^2)$).  
There is no change on the actual value of $\chi^2$ but its expression is 
easier (though not faster!) to evaluate.}
(the estimated uncertainty $\sigma_j$ of the measurement).  
$\Delta v$, together with $\frac{\partial 
v}{\partial p_k}$ ($k = 1$ to 5, with $p_1 \equiv \alpha_0^*, p_2 \equiv \delta,
p_3 \equiv \varpi, p_4 \equiv \mu_{\alpha^*}, p_5 \equiv \mu_\delta$), the
original astrometric parameters 
as well as $\rho, \sigma_{F_j}$ and $\sigma_{N_j}$ are given in the IAD file. 
In order to evaluate $\chi^2$ (Eq.~\ref{chi2}) for an orbital model, 
expressions for the partial derivatives of $v$ with respect to the 
orbital parameters are required. They can be expressed as a 
function of the partial derivatives of $v$ with respect to $\alpha_0^*$ and 
$\delta_0$ as follows (Hipparcos catalogue, Vol.~3, Eq.~$17.15$): 
\[
\frac{\partial v}{\partial o}=\frac{\partial v}{\partial 
\alpha_0^*}\frac{\partial \xi}{\partial o}+\frac{\partial 
v}{\partial \delta_0}\frac{\partial \eta}{\partial o}
\]
where $o$ is any orbital parameter and
\begin{equation}
\begin{array}{l}
\xi=\alpha_0^*+\mu_{\alpha^*}(t-t_0)+R P_{\alpha}\varpi+y,\\ 
\eta=\delta_0+\mu_{\delta}(t-t_0)+R P_{\delta}\varpi+x.
\end{array}\label{Eq:xieta}
\end{equation}
In the above expression, $\xi$ and $\eta$ represent the Cartesian 
coordinates of the observed component on the plane tangent to the 
sky at the position ($\alpha_0, \delta_0$). They combine the 
displacements due to the proper motion, the orbital motion and the parallax. 
$R$ is the radius vector of the Earth's orbit in A.U. at time $t$.  $P_{\alpha}$
and $P_{\delta}$, the parallax factors, are given by \cite*{PrDoSt}: 
\begin{eqnarray*} 
P_{\alpha}&=&\cos\epsilon\cos\alpha\sin\sun -\sin\alpha\sin\sun\\
P_{\delta}&=&(\sin\epsilon\cos\delta-\cos\epsilon\sin\alpha\sin\delta)\sin\sun\\
&&{} -\cos \alpha \sin \delta \cos \sun
\end{eqnarray*}
where $\sun$ and $\epsilon$ are respectively the longitude of the Sun and 
the obliquity of the ecliptic, both at time $t$.The variables $x$ and $y$ 
describe the apparent orbit (i.e., projected on the plane orthogonal to the 
line of sight) of the observable component around the center of mass of the 
system.  They are usually expressed in terms of the Thiele-Innes constants $A, 
B, F, G$ of the photocentric orbit as (\cite{DoSt})
\begin{equation}
\begin{array}{l}
x=AX + FY\\
y=BX + GY
\end{array}
\label{Eq:xy}
\end{equation} 
with 
\begin{eqnarray*} 
X &=& \cos E -e \label{Eq:CXCY}\\ 
Y &=& \sqrt{1 - e^2} \sin E,
\end{eqnarray*}
where $E$ is the eccentric anomaly and $(X,Y)$ are the coordinates in the true
orbit.  It should be noted that $(x,y)$ and $(X,Y)$ are referred to a Cartesian
system with $x$ pointing towards the North (\cite{DoSt}), contrary to $(\xi,
\eta)$ where $\xi$ points towards increasing right ascensions.
For all the systems considered in the present paper, the magnitude difference
between the two components is larger than the Hipparcos detection threshold
(since the companion to CPRS is a cool white dwarf; Sect.~\ref{Sect:masses} and
\cite{Jorissen-1998}).  Thus, one can safely assume that
there is no light coming from the secondary. Hence,  the orbit of the
photocenter of the primary component 
is the same as the absolute orbit of the primary around the center of mass of
the system. 

At this point, it is important to realize that the Hipparcos 5-parameter 
solution, that is used as a starting point for the new 12-parameter solution,
is in fact equivalent to a 12-parameter solution where the semi-major axis 
of the orbit ($a_0$) is null. It is therefore enough to consider in 
Eq.~\ref{chi2} the correction term relative to the semi-major axis of 
the orbit (and in fact $\Delta a_0 = a_0$, since the initial value of $a_0$ 
is null). The other orbital parameters $i, \omega, \Omega, e, P$ and $T$ 
enter Eq.~\ref{chi2} only through the partial derivatives $\frac{\partial 
\xi}{\partial a_0}$ and $\frac{\partial \eta}{\partial a_0}$ [equal, 
respectively, to $\frac{\partial y}{\partial a_0}$ and 
$\frac{\partial x}{\partial a_0}$ according to Eq.~\ref{Eq:xieta}]   
entering $\frac{\partial v}{\partial 
a_0}$. These other orbital parameters do not require explicit 
correction terms in Eq.~\ref{chi2} (and their starting values would be 
ill-defined anyway).  Hence, the orbital solution is the one 
minimizing 
\begin{equation}
\chi^2=\Xi^{\mbox{t}}V^{-1}\Xi 
\label{Eq:newchi2}
\end{equation}
where
\begin{eqnarray}
\label{Eq:Capitalchi}
\Xi&=&\Delta v-\sum_{k=1}^5\frac{\partial v}{\partial p_k}\Delta 
p_k - (\frac{\partial v}{\partial p_1}\frac{\partial \xi}{\partial 
a_0}+\frac{\partial v}{\partial p_2}\frac{\partial \eta}{\partial 
a_0}) a_0
\end{eqnarray}
is a vector of dimension N (N is the number of observations), and $\xi$ and 
$\eta$ are functions of $a_0, i, \omega, \Omega, e, 
P$ and $T$, and $\frac{\partial\xi}{\partial a_0} \equiv \frac{\partial
y}{\partial a_0}, \frac{\partial\eta}{\partial a_0} \equiv \frac{\partial
x}{\partial a_0}$.  Thus, the 12 parameters enter the evaluation of 
$\chi^2$ although there are only six correction terms subtracted 
from $\Delta v$.

Our experience has shown that, except for very special cases (i.e., 
parallaxes and semi-major axes larger than about 20~mas, 
orbital periods significantly different from 1 year but smaller than 3 years;
one example is HIP~50805 in Table \ref{Tab:criteria}), astrometric orbits 
could not be derived from the IAD without an {\it a priori} knowledge of 
some of the orbital elements, for instance the spectroscopic ones ($e$, $P$
and $T$).  However, it appeared  that fixing $\omega$ at the value derived 
from the spectroscopic orbit often led to orbital inclinations $i$ 
unrealistically close to zero for spectroscopic binaries with radial 
velocity variations.  Leaving the parameter $\omega$  free removes 
this difficulty, and offers moreover a way to check the consistency 
of the astrometric solution, since the astrometric $\omega$  
should  be consistent with its spectroscopic value.

As far as outliers are concerned, we almost always keep the same data set 
as that used by FAST and/or NDAC (Vol. 3, Sect. 17.6;\cite*{Hipparcos}). 
IAD that were not considered by FAST or NDAC (i.e., with the IA2 field set 
to `f' or `n') were thus not included in our re-processing either.  In a few 
cases, we noticed that because of the orbital contribution, some observations
yield residuals larger than $3\sigma$ of the residuals.  In these cases, these 
observations were removed and the fit re-iterated.  We never had to iterate 
more than twice to remove all outliers.

\subsubsection{Forcing parallaxes to be positive}

One of the five astrometric parameters entering $\chi^2$ in Eq.~\ref{chi2} 
is the parallax $\varpi$. With no other prescriptions as those described 
in Sect.~\ref{Sect:IAD:OF}, the minimization process may very well end up with
a negative parallax. Indeed, negative parallaxes are not rare in the 
Hipparcos and Tycho catalogues.

Parallaxes cannot only be seen as inverse distances 
(which are defined positive)
but also as the semi-major axis of the parallactic ellipse (see
Eq.~\ref{Eq:xieta}). The direction of motion along the parallactic ellipse is
of course imposed by the annual revolution of the Earth around the Sun,
regardless of the actual dimension of the parallactic ellipse or of the
observational uncertainties. 
In that sense, the parallactic ellipse is {\it oriented}, and negative
parallaxes can be seen as corresponding to a parallactic ellipse 
covered in the wrong direction.  That constraint being of physical nature, one
should seek to fulfill it. In this section, we present a method which always
delivers positive parallaxes. This method is especially useful for stars like
those Mira variables or carbon stars that came out with 
large negative parallaxes in the
Hipparcos Catalogue. In those cases, forcing the parallax to be positive
has a strong impact on the derived proper motion (Pourbaix et al., in
preparation), which may be supposed to be better determined with a physically-
sound model yielding positive parallaxes.  However,
the major drawback of the method is that the errors on the parallax do
no longer follow a normal distribution. Therefore, the use of the parallaxes 
provided by this paper for
e.g., luminosity calibrations should be done with care to avoid
biases.

In order to force the parallax to be positive -- and at the same time 
avoiding the difficulties inherent to any constrained minimization techniques
-- one may replace the constrained variable (the parallax $\varpi$) by an 
unconstrained one\footnote{We are indebted to A.~Albert for suggesting this 
trick}. An appropriate choice appears to be 
\begin{equation}
\label{Eq:logit}
\varpi'=\log \varpi
\end{equation}
which is $C^{\infty}$ between $]0,+\infty[$ 
and $\mathbb{R}$.   The variable $\varpi'$ is in fact
equivalent 
to the distance modulus $m-M$ since
\begin{equation}
\label{Eq:distancemod}
m-M = -5 - 5 \varpi'.
\end{equation}
The variable  $\varpi'$ is thus used in the minimization 
process instead of 
$\varpi$ as one of the $p_k$  parameters entering Eq.~\ref{chi2} [with 
$\frac{\partial v}{\partial \varpi'} = \frac{\partial v}{\partial 
\varpi} \varpi \ln 10$].  At the end of the minimization, 
Eq.~\ref{Eq:logit} is reversed and the parallax $\varpi$ is derived from 
$\varpi'$.  For any real number $\varpi'$, $\varpi$ necessarily lies in 
between 0 and $\infty$.  Such a substitution is legitimate, since 
the parallax is not a directly measured quantity, but rather one among many 
parameters used in a model fit to the observations. Moreover, since
maximum-likelihood estimators\footnote{$\chi^2$ is a
maximum-likelihood estimator provided that the measurement errors
are normally distributed} enjoy the {\it invariance property} (see
e.g., \cite{InThSt}, p. 284), $\varpi$- and
$\varpi'$-fitting must yield identical results for those cases where
$\varpi$-fitting yielded a positive parallax (since the 
$\varpi'=\log \varpi$ transform may be used when $\varpi > 0$ 
to extract $\varpi'$). 

Most of the objects considered in the present paper have large parallaxes that
would have come out positive by  a direct fit of
$\varpi$ anyway. 
Thus, in the present case, the fitting of $\varpi'$ (instead of
$\varpi$) does not
represent so much of an improvement. Nevertheless, the procedure of 
$\varpi'$-fitting has been introduced here for the sake of generality.

The price to pay is, however, that the errors on $\varpi$ do not any 
longer follow a normal distribution. Moreover, the confidence interval of 
the parallax  is no more 
symmetric. It has been estimated by the following expression:
\begin{equation}
\begin{array}{l} 
\hat{\varpi}'-\sigma\le\varpi'\le\hat{\varpi}'+\sigma\\ 
10^{\hat{\varpi}'-\sigma} \le
\varpi\le 10^{\hat{\varpi}'+\sigma}
\end{array}
\label{Eq:piprimerr}
\end{equation}
where $\hat{\varpi}'$ is the value resulting from the $\chi^2$-
minimization, and $\sigma$ its estimated uncertainty. It should
be noted that the errors 
on $\varpi'$ do not follow a normal distribution either, 
since the normality is only guaranteed for `linear models' (\cite{InThSt}) 
and  $\Xi$ (Eq.~\ref{Eq:newchi2}) 
does not depend linearly upon $\varpi'$. Therefore, the confidence
interval corresponding to a given probability level is in general
not symmetric around $\varpi'$. However, to ease the computations,
the uncertainty $\sigma$ on $\varpi'$ will be computed from the 
inverse of the Fisher information matrix at the point minimizing $\chi^2$ 
(see Sect.~\ref{Sect:minimiz}). This procedure implicitly assumes
that the model is linear in the vicinity of the minimum,  
so that the adopted confidence interval is in fact one that is symmetric 
around $\varpi'$.

The second expression in Eq.~\ref{Eq:piprimerr} clearly shows that 
the parallax is positive everywhere in the confidence interval, which 
would not be guaranteed in a constrained minimization of $\chi^2$.  That
important property illustrates the superiority of this approach with 
respect to the constrained minimization.

\subsubsection{$\chi^2$ minimization}
\label{Sect:minimiz}

If the $\chi^2$ expressed by Eq.~\ref{Eq:newchi2} were a quadratic expression 
of the unknown parameters $p_k$, its unique minimum could be found 
from the solution of a set of linear equations. However, the parameters  
$i$, $\omega$ and $\Omega$ enter $\chi^2$ in a highly non-linear way, so that 
the function expressing $\chi^2$ in the 9-parameter space may have several 
local minima, and finding its global minimum is a much more arduous task. 

Faced with such situations, one of us (DP) has already successfully worked out
global optimization techniques such as {\it simulated annealing}
(\cite{Pourbaix-1994,Pourbaix-1998:a}).  Practical details about the 
implementation of the method to minimize the objective function $\chi^2$
[Eq.~\ref{Eq:newchi2}] in the working space $\mathbb{R}^9$ may be found in 
\cite*{Pourbaix-1998:b}.  Simulated annealing being a heuristic method, one
can only prove its convergence to the global minimum after an infinite time
(which we cannot afford).  We thus stop the procedure after a finite time.
In order to nevertheless have a good chance to obtain (a neighborhood of) 
the global minimum, we repeat 40 times this highly non deterministic 
minimization process.  The best solution ever met (i.e., the one leading to 
the smallest $\chi^2$ value) is finally adopted.  Once (a neighborhood of) 
the global minimum is thus obtained, it is tuned with the BFGS quasi-Newton 
algorithm (\cite{NuMeUnOpNoEq}).

Unlike the Levenberg-Marquardt (\cite{Marquardt-1963}) minimization algorithm,
BFGS does not return the covariance matrix of the model parameters.  The 
inverse of the Fisher information matrix at the minimum is therefore used as 
the best estimate of that covariance matrix (\cite{Pourbaix-1994}).

The whole procedure has been applied separately on the data from the FAST
consortium only, from NDAC only and from both combined, thus resulting in 
three different solutions, hopefully consistent with each other. 

In a few instances, the solution obtained from the combined FAST+NDAC data set
turns out to be very close to either the FAST or NDAC solution, but FAST and NDAC
taken separately yield rather different solutions. That situation probably
reflects the very different weights attributed to the two data sets for that
particular object in the merging process applied to produce the output catalogue.
For our analysis we always keep the covariance matrices of the observations
as they are given in the electronic version of the catalogue.

As pointed out by an anonymous referee, in the case where
all the Campbell elements ($a_0$, $i$, $\omega$ and $\Omega$) are extracted from
a fit to the astrometric data,
they can advantageously be replaced by the Thiele-Innes elements ($A$, $B$, 
$F$ and $G$) so that $\chi^2$ becomes a quadratic function of the model 
parameters. The minimum of $\chi^2$ can then be found analytically and no 
minimization (neither global nor local) technique is needed.  For the sake of 
generality, we nevertheless use the Campbell set (and thus the minimization
scheme) because this more general scheme allows, if necessary, to easily
incorporate external constraints (like for instance the knowledge of $i$ for
eclipsing binaries, or $\omega$ from the spectroscopic orbit; see, however, the
comment about fixing $\omega$ after Eq.~\ref{Eq:Capitalchi} in
Sect.~\ref{Sect:IAD:OF}). 
Such additional constraints would be much more difficult
to impose through the Thiele-Innes elements.

\subsubsection{F-test: 5-parameter vs orbital model} 
\label{Sect:Ftest}

The introduction of more free parameters in the orbital model as 
compared to the single star solution  necessarily leads to a reduction of 
the objective function. To evaluate whether this reduction  is statistically
significant -- or, equivalently, whether the orbital solution represents a 
significant improvement over the 5-parameter Hipparcos solution -- requires 
the use of an F-test. 

The method used here is inspired from the test devised by \cite*{Lucy-1971}.  
If $\chi^2_{\rm Hip}$ and $\chi^2_{\rm orb}$ denote the residuals 
for the Hipparcos 5-parameter model and for the orbital model (with 
9 free parameters), 
respectively, the efficiency of the additional 4 parameters in reducing 
$\chi^2_{\rm orb}$ below $\chi^2_{\rm Hip}$ may be measured by 
the ratio:
\begin{equation}
\label{Eq:Ftest}
F =\frac{N-9}{4}
         \frac{\chi^2_{\rm Hip} - \chi^2_{\rm orb}}{\chi^2_{\rm orb}}, 
\end{equation}
where $N$ is the number of available measurements.

If the hypothesis that there is no orbital motion (i.e., $a_0 = 0$ in
Eq.~\ref{Eq:newchi2})  is correct, then it may be shown (\cite{DaReErAnPhSc}) 
that $F$ follows a Snedecor
$F_{\nu_1,\nu_2}$ distribution with $\nu_1 = 4$ and $\nu_2 = N-9$
degrees of freedom. Thus, if Eq.~\ref{Eq:Ftest} yields $F = \hat{F}$,
then, on the assumption that there is no orbital motion, the
probability that $F$ could have exceeded $\hat{F}$ is 
\begin{equation}
\label{Eq:F}
\alpha = \mbox{Prob}(F > \hat{F}) \equiv Q(\hat{F}|\nu_1,\nu_2).  
\end{equation}
In other words, $\alpha$ is the first risk error of rejecting the null 
hypothesis that the orbital and 5-parameter models are identical while it 
is actually true.

The residuals given in the IAD files always relate to a 5-parameter
solution, even when a more sophisticated model (the so-called `acceleration'
7- or 9-parameter models, or even orbital model) was published in the 
Hipparcos catalogue (see Table~\ref{Tab:criteria}).  For those cases, the 
$\alpha$ value listed in Table~\ref{Tab:criteria} is always close to zero, 
although it does not really characterize the improvement of the orbital 
solution with respect to the solution retained in the Hipparcos catalogue 
(which goes already beyond a 5-parameter model).

\subsubsection{Astrophysical consistency of the orbital solution} 
\label{Sect:test}

The minimization process will yield a solution in all cases, but 
that solution may not be astrophysically relevant. A statistical 
check of the significance of the orbital solution, based on 
the $F$-test,  has been presented in Sect.~\ref{Sect:Ftest}.  In this 
section, two criteria testing the validity of the orbital solution on 
astrophysical grounds are presented.

The first test is based upon the identity 
\begin{equation}
\label{Eq:check1}
\frac{a_0\sin i}{\varpi} = \frac{P\;  K_1\; \sqrt{1-e^2}}{2\pi},
\end{equation}
where $K_1$ is the radial-velocity semi-amplitude of the visible component. 
The left-hand 
side of the above identity entirely depends upon astrometric 
parameters, whereas its right-hand side contains only  parameters 
derived spectroscopically.

This test has the advantage of being totally independent of any 
assumptions. However, it involves the orbital inclination which is 
not always very accurately determined (see Sect.~\ref{Sect:checks}), so 
that the above identity may not always be very constraining considering 
the often large uncertainty on $i$. 

A somewhat more constraining identity to assess the astrophysical
plausibility of the computed astrometric orbit is the following:
\begin{equation}
\label{Eq:a0varpi}
\frac{a_0}{\varpi} = P^{2/3} \frac{M_2}{(M_1 + M_2)^{2/3}},
\end{equation}
where $M_1$ and $M_2$ are the masses of the visible and invisible 
components, respectively.
However, it relies upon assumptions regarding the stellar masses. 
In some cases (e.g., dwarf barium stars), the mass of the 
observable component may be estimated directly from the 
spectroscopically-derived gravity and from an estimate of the 
stellar radius from the spectral type (\cite{North-1999:a}). 
For the other samples, an average mass derived from a statistical 
analysis of the spectroscopic mass function (\cite{Jorissen-1998}) is 
adopted. For  all the samples considered here, the unseen 
component is almost certainly a white dwarf (the only possible
exception being the Tc-poor S star HIP~99312 = HD~191589), whose mass may be 
taken as $0.62\pm0.04$~\Msun\  (\cite{Jorissen-1998}). 

Because of the assumptions involved, this test is  only used as a guide to
identify astrophysically-unplausible solutions.
It turns out that such cases are generally those with large error bars or with
inconsistent N, F and A solutions, thus providing further arguments not to retain
those solutions. In very few cases (HIP~36042, 53763 and 60299), valid 
data yielded solutions not consistent with Eq.~\ref{Eq:a0varpi}.  Those cases
were nevertheless kept in our final list.

\subsection{Transit Data} 

Unlike the IAD, TD are only available for a small subset of the Hipparcos
catalogue, e.g., for those stars that were known to be (or suspected of 
being) double or multiple systems at the time of the data reduction by the 
Hipparcos consortia.  TD are a by-product of (or, more precisely, an input 
for) the multiple-star processing by the NDAC consortium.  Another difference
with respect to the IAD concerns photometry.  Whereas IAD contain astrometric
information only, the brightness of the (different) {\em observable} 
component(s) of the system can be retrieved from the TD.

In the most general case (\cite{Quist-1999}), each entry in the TD file
corresponds to
five numbers $b_1$, \dots, $b_5$ which represent the coefficients of the first
terms in the Fourier series modeling the observed signal as modulated by the
detector grid:
\begin{eqnarray*}
b_1&=&\sum_iI_i\\
b_2&=&\tilde M\sum_iI_i\cos\phi_i\\
b_3&=&-\tilde M\sum_iI_i\sin\phi_i\\
b_4&=&\tilde N\sum_iI_i\cos(2\phi_i)\\
b_5&=&-\tilde N\sum_iI_i\sin(2\phi_i)\\
\end{eqnarray*}
where $I_i$ is the intensity of the $i$-th component of the system, and the
phase $\phi_i$ corresponds to its abscissa along the reference great circle.
$\tilde M=0.7100$ and $\tilde N=0.2485$ are the adopted reference
values for the modulation coefficients of the first and second
harmonics (Vol.~1, Sect.~2.9; 
\cite{Hipparcos}). In terms of the Cartesian coordinates $(\xi_i, \eta_i)$ 
(see Eq.~\ref{Eq:xieta})
of component $i$ in the plane tangent to the sky at the reference point
specified in the TD file, the 
phase $\phi_i$ writes
\begin{equation}
\phi_i=f_x\xi_i+f_y\eta_i+f_p(\varpi - \varpi_{\rm ref}). \label{Eq:phiComputed}
\end{equation}
The reference point is assigned an arbitrary parallax $\varpi_{\rm ref}$ and
proper motion, as given in the TD file. The phase derivatives $f_x, f_y$ and
$f_p$ with respect to $\xi, \eta$ and $\varpi$ are also provided by the TD file.

For the SB1 systems we are interested in, the situation simplifies a lot since
$I_2$ may be taken equal to 0.  The above system of equations is rank-deficient. 
From the second and third equations, one can rewrite:
\begin{equation}
\stackrel{o}{\phi}=\arg(b_2,-b_3) \label{Eq:phiObserved}
\end{equation}
and also express the uncertainty on $\phi$ as
\[
\sigma_{\phi}^2=\left(\frac{\partial \stackrel{o}{\phi}}{\partial
  b_2}\right)^2\sigma_{b_2}^2+\left(\frac{\partial \stackrel
{o}{\phi}}{\partial b_3}\right)^2\sigma_{b_3}^2. 
\]

With the above definitions, the objective function whose minimum is sought is
given by:
\begin{equation}
D=\frac{1}{N}\sum_{k=1}^N\sigma_{\phi_k}^{-2}(\stackrel{o}{\phi_k}-\phi_k)^2
\label{eq:ObjFuncTD}
\end{equation}
where $N$ is the number of TD. In the above expression, $\stackrel{o}{\phi_k}$
is the observed phase at time $t_k$ as derived from Eq.~\ref{Eq:phiObserved},
whereas $\phi_k$ is the phase computed from Eqs.~\ref{Eq:xieta} and
\ref{Eq:phiComputed} (noting that $f_x R P_\alpha + f_y R P_\delta \equiv f_p$)
for a given set of astrometric and orbital parameters.

The remaining of the method follows the same steps as described in relation 
with the IAD, i.e. global and local optimization, positiveness of the 
parallax, \dots

\section{Results}

Table~\ref{Tab:criteria} lists the various samples that have been considered, 
namely dwarf barium and subgiant CH stars (\cite{McClure-1997:a}; North, priv. 
comm.), mild and strong barium stars
(\cite{Jorissen-1998}), Tc-poor S stars (\cite{Jorissen-1998}) and CH stars 
(\cite{McClure-1990}). Their spectroscopic orbital parameters have been taken
from the reference quoted. 

Table~\ref{Tab:criteria} also provides various parameters that allow either to 
assess the quality of the derived orbital solution 
or to identify the reason why a reliable orbital solution could not be derived
for some of the spectroscopic binaries considered.  
The following parameters potentially control our ability to derive an
astrometric orbit from the IAD, with favorable circumstances being mentioned
between the parentheses: the parallax (column~3; large parallax), the number 
of available IAD measurements (FAST+NDAC; column~4; large number of
measurements), the ecliptic latitude
(column~5; this parameter may play a role since it controls 
how different the orientations of the reference great circles are; favorable
cases have absolute values larger than 47\degr), the orbital period 
(column~6; in the range 1 -- 3~yr to ensure a good sampling of the orbit), the
orbital eccentricity (column~7; low eccentricity).

\begin{table*}[htb]
\caption[]{\label{Tab:criteria}
The various samples of CPRS that have been
considered in the present work, along with 
various parameters characterizing the quality of the orbital solution derived
from the IAD (see text for details). $\varpi_{\rm HIP}$ is the parallax from 
the Hipparcos catalogue, $N_{\rm IAD}$ is the number of IAD available for the
star considered, $\beta$ is the ecliptic latitude, $T$ is the time of periastron
passage (or of maximum velocity for circular orbits). Sol$_{\rm HIP}$ is the
number of parameters used in the Hipparcos
solution. 
It may be one of the following types: `5' = 5-parameter
solution, `7' 
= 7-parameter solution, i.e., including a linear  acceleration term 
in the proper motion (see Sect. 2.3, Vol.~1 of \cite{Hipparcos}), `9' 
= 9-parameter solution, i.e., including a quadratic acceleration 
term in the proper motion (`7'- and `9'-entries are listed in the `G' section
of the {\it Double and Multiple Star Annex} -- DMSA -- of the Hipparcos and
Tycho Catalogues), `X' = stochastic solution (described in DMSA/X),
and `O' = orbital solution given in the DMSA/O. The column labeled 
`Transit?' indicates whether TD are also available for the corresponding star.
The column labeled $\chi^2/(N_{\rm IAD}-9)$ lists the 
unit-weight variance for the orbital-parameter model. 
$\alpha$ is the first risk error of rejecting the null hypothesis that the 
orbital and single star models are identical while it is actually true 
(Eq.~\ref{Eq:F}).  The expected value of the orbital separation $a_0/\varpi$
is computed from Eq.~(\ref{Eq:a0varpi}).  The $1\sigma$ confidence interval 
is given next to the computed astrometric $a_0/\varpi$ value. Column `Rem'
indicates whether the orbital solution derived 
from the IAD has been accepted. An
asterisk in column `Rem' refers to a note at the bottom of the Table}
\renewcommand{\arraystretch}{0.9}
\renewcommand{\tabcolsep}{4pt}
\begin{scriptsize}
\begin{tabular}{llllrlllcclllll}\hline
HIP&HD/DM&$\varpi_{\rm HIP}$&$N_{IAD}$&$\beta$&$P$&$e$&\multicolumn{1}{c}{$T$}
&Sol$_{\rm
HIP}$&Transit?&$\frac{\chi^2}{(N_{\rm
IAD}-9)}$&$\alpha$&\multicolumn{2}{c}{$a_0/\varpi$}&Rem\\
\cline{13-14}&&(mas)&&(\degr)&(d)&\multicolumn{3}{c}{(JD-
2\ts400\ts000)}&&&(\%)&astrom.&expected&\\\hline
\multicolumn{2}{l}{\bf CH stars}
\medskip\\
168&224959&1.95$\pm$1.34&62&-2.8&1273.0&0.179&46064&5&y&1.23&36&1.6$\begin{array}{l}
5.1\\0.41\end{array}$&1.08$\pm$0.53& \\
4252&5223&1.12$\pm$1.17&50&16.8&755.2&0&45535.6&5&&0.69&
0&1e+09$\begin{array}{l}+\infty\\0\end{array}$&0.76$\pm$0.38& \\
22403&30443&1.63$\pm$1.7&28&12.5&2954.0&0&46306&5&y&0.90&
9&24$\begin{array}{l}67\\3.2\end{array}$&1.89$\pm$0.94& \\
53763&$+42^\circ2173$&5.29$\pm$1.47&70&32.2&328.3&0&46542.1&5&&0.84&
3&1.8$\begin{array}{l}8.3\\0.3\end{array}$&0.44$\pm$0.22&Accepted,*\\
62827&$+8^\circ2654$A&5.3$\pm$2.15&40&11.8&571.1&0&46467.8&5&&1.10&89&0.42$\begin{array}{l}1.5\\0\end{array}$&0.63$\pm$0.31&*\\
102706&198269&3.16$\pm$1.11&68&34.2&1295.0&0.094&46358&5&&1.18&34&2.5$\begin{array}{
l}5.6\\0.78\end{array}$&1.09$\pm$0.54& \\
104486&201626&4.93$\pm$0.84&100&40.6&1465.0&0.103&45970&7&&1.35&
0&1.5$\begin{array}{l}2.3\\0.84\end{array}$&1.18$\pm$0.58&Accepted\\
108953&209621&1.47$\pm$1.3&46&30.6&407.4&0&45858.3&5&&1.02&94&0.8
$\begin{array}{l}9.3\\0\end{array}$&0.50$\pm$0.25& 
\medskip\\
\multicolumn{2}{l}{\bf Dwarf Ba stars}
\medskip\\
8647&11377&6.38$\pm$1.14&42&-26.3&4140.0&0.16&45240&5&y&0.95&34&9.5$\begin{array}{l}
27\\0\end{array}$&2.20$\pm$0.97&\\
32894&50264&14.11$\pm$1.96&78&-52.2&912.4&0.098&46791&X&&1.05&
0&0.84$\begin{array}{l}1\\0.7\end{array}$&0.89$\pm$0.46&Accepted \\
49166&87080&7.9$\pm$1.39&90&-42.2&273.4&0.177&48373&5&&1.03&34&0.47$\begin{array}{l}
0.85\\0.2\end{array}$&0.39$\pm$0.19&Accepted\\
50805&89948&23.42$\pm$0.93&60&-36.6&667.8&0.117&46918&O&y&0.82&
0&0.67$\begin{array}{l}0.73\\0.6\end{array}$&0.67$\pm$0.30&Accepted\\
60299&107574&5.02$\pm$1.06&50&-14.7&1350.0&0.081&46342&9&&0.93&
0&1.8$\begin{array}{l}2.5\\1.2\end{array}$&0.83$\pm$0.26&Accepted\\
62409&+17$^\circ2537$&8.2$\pm$1.28&58&20.1&1796.0&0.14&46291&5&&1.29&25
&0.87$\begin{array}{l}1.7\\0.28\end{array}$&1.34$\pm$0.65&NDAC accepted\\
69176&123585&8.75$\pm$1.39&54&-29.3&457.8&0.062&48207&5&&1.24&
0&0.47$\begin{array}{l}0.69\\0.31\end{array}$&0.44$\pm$0.16&Accepted \\
71058&127392&10.63$\pm$1.7&46&-15.4&1498.7&0.071&47070&5&&0.85&73&0.56$\begin{array}
{l}1.4\\0\end{array}$&1.21$\pm$0.61& \\
104785&202020&9.53$\pm$1.5&52&6.1&2064.0&0.08&47122&5&&1.16&14&1.4$\begin{array}{l}3
.2\\0.14\end{array}$&1.51$\pm$0.77& \\
105969&204613&16.61$\pm$1.78&80&64.8&878.0&0.13&47479&X&&1.02&
0&0.75$\begin{array}{l}0.84\\0.66\end{array}$&0.77$\pm$0.33&Accepted \\
107818&207585&7.53$\pm$1.51&52&-10.5&670.6&0.03&47319&5&&0.82&
7&0.39$\begin{array}{l}0.65\\0.2\end{array}$&0.58$\pm$0.22&Accepted\\
116233&221531&8.83$\pm$1.21&56&-8.3&1416.0&0.165&47157&7&&0.84&
0&1.2$\begin{array}{l}1.5\\0.89\end{array}$&0.97$\pm$0.37&Accepted \\
118266&224621&6.95$\pm$1.45&80&-32.6&307.8&0.048&49345&5&&1.00&45&0.27$\begin{array}
{l}0.55\\0.081\end{array}$&0.40$\pm$0.19&\\
\hline
\end{tabular}
\end{scriptsize}

Remarks:\\
HIP 53763 \& 62827: large uncertainty on $\varpi$ in the 9-parameter
solution (see Table~\protect\ref{Tab:orbits}), since the orbital
period is close to 1~yr.\\
\end{table*}

\addtocounter{table}{-1}
\begin{table*}[htb]
\caption{(Continued.)}
\renewcommand{\tabcolsep}{4pt}
\begin{scriptsize}
\begin{tabular}{llllrlllcclllll}\hline
HIP&HD/DM&$\varpi_{\rm HIP}$&$N_{IAD}$&$\beta$&$P$&$e$&\multicolumn{1}{c}{$T$}
&Sol$_{\rm
HIP}$&Transit?&$\frac{\chi^2}{(N_{\rm
IAD}-9)}$&$\alpha$&\multicolumn{2}{c}{$a_0/\varpi$}&Rem\\
\cline{13-14}&&(mas)&&(\degr)&(d)&\multicolumn{3}{c}{(JD-
2\ts400\ts000)}&&&(\%)&astrom.&expected&\\\hline
\multicolumn{2}{l}{\bf Mild Ba stars}
\medskip\\
19816&26886&2.74$\pm$1.05&48&-21.9&1263.2&0.395&48952.12&5&&0.55&
1&5.2$\begin{array}{l}15\\0.61\end{array}$&0.77$\pm$0.23& \\
20102&27271&6.01$\pm$1.13&28&-18.6&1693.8&0.217&47104.38&5&&0.66&46
&1.8$\begin{array}{l}5.6\\0\end{array}$&0.93$\pm$0.28& \\
26695&288174&2.89$\pm$1.32&42&-21.3&1824.3&0.194&47157.62&5&&0.80&66&5.1$\begin{array}{
l}15\\0.74\end{array}$&0.98$\pm$0.29& \\
32831&49841&-1.05$\pm$1.44&44&-17.2&897.1&0.161&48339.71&5&y&1.15&44&1.8e+09$\begin{array}{l}+\infty\\0\end{array}$&0.61$\pm$0.18& \\
34143&53199&3.65$\pm$1.3&44&-9.3&7500.0&0.212&41116.2&5&y&0.99&72&26$\begin{array}{l}1
10\\0\end{array}$&2.51$\pm$0.75& \\
35935&58121&2.82$\pm$0.95&64&-15.7&1214.3&0.14&46811.21&5&&1.12&53&0.98$\begin{array}{l}2.4\\0.22\end{array}$&0.75$\pm$0.22& \\
36042&58368&2.36$\pm$0.97&56&-14.3&672.7&0.221&45617&5&&1.00&
1&2$\begin{array}{l}3.8\\0.97\end{array}$&0.50$\pm$0.15&Accepted\\
36613&59852&-1.86$\pm$1.24&60&-25.8&3463.9&0.152&46841.03&5&&1.25&100&6.2e+51$\begin{array}{l}+\infty\\0\end{array}$&1.50$\pm$0.45& \\
43527&$-14^\circ2678$&0.41$\pm$1.4&56&-30.9&3470.5&0.217&48828.06&5&y&0.84&59&49$
\begin{array}{l}3000\\0.14\end{array}$&1.50$\pm$0.45& \\
44464&77247&2.86$\pm$0.97&46&34.6&80.5&0.0871&48953&5&y&1.05&93&0.39$\begin{array}{l
}1.3\\0\end{array}$&0.12$\pm$0.04& \\
51533&91208&4$\pm$0.95&48&-24.4&1754.0&0.171&45628.36&5&&1.45&72&3.5$\begin{array}{l}8.
7\\0.41\end{array}$&0.95$\pm$0.28& \\
53717&95193&2.3$\pm$1.03&44&-19.0&1653.7&0.135&46083.62&5&&1.15&89&5.3$\begin{array}{l}
22\\0\end{array}$&0.92$\pm$0.27& \\
73007&131670&2.33$\pm$1.22&34&9.0&2929.7&0.162&46405.11&5&&1.17&45&21$\begin{array}{l}1
30\\0\end{array}$&1.34$\pm$0.40& \\
76425&139195&13.89$\pm$0.7&38&28.5&5324.0&0.345&44090&5&&0.90&61&25$\begin{array}{l}
47\\5.3\end{array}$&2.00$\pm$0.59& \\
78681&143899&3.6$\pm$1.29&36&1.2&1461.6&0.194&46243.43&5&&1.65&93&1.2$\begin{array}{l}4
.6\\0\end{array}$&0.84$\pm$0.25& \\
94785&180622&3.37$\pm$1.04&54&22.5&4049.2&0.061&50534.41&5&&1.03&52&26$\begin{array}{l}
67\\3.3\end{array}$&1.66$\pm$0.50& \\
103263&199394&6.33$\pm$0.63&84&59.6&4382.6&0&50719.34&5&&0.81&76&4$\begin{array}{l}11\\
0\end{array}$&1.75$\pm$0.53& \\
103722&200063&0.73$\pm$1.02&48&17.2&1735.5&0.073&47744.64&5&&1.00&
3&14$\begin{array}{l}96\\0.25\end{array}$&0.95$\pm$0.28& \\
104732&202109&21.62$\pm$0.63&98&43.7&6489.0&0.22&40712&7&&0.71&
0&6$\begin{array}{l}10\\1.8\end{array}$&2.28$\pm$0.68& \\
105881&204075&8.19$\pm$0.9&54&-7.0&2378.2&0.2821&45996&5&y&0.92&
6&1$\begin{array}{l}1.7\\0.43\end{array}$&1.17$\pm$0.35&Accepted\\
106306&205011&6.31$\pm$0.68&68&36.2&2836.8&0.2418&46753.59&5&&0.71&
3&1.7$\begin{array}{l}+\infty\\0\end{array}$&1.31$\pm$0.39& \\
109747&210946&3.42$\pm$1.14&36&11.7&1529.5&0.126&46578.18&5&&0.89&14&2.2$\begin{array}{
l}4.9\\0.68\end{array}$&0.87$\pm$0.26& \\
112821&216219&10.74$\pm$0.93&56&23.3&4098.0&0.101&44824.92&5&&0.85&25&3$\begin{array}{l
}6.8\\0\end{array}$&1.96$\pm$0.74& *\\
117607&223617&4.61$\pm$0.95&62&2.9&1293.7&0.061&47276.68&7&&0.75&
0&0.87$\begin{array}{l}1.5\\0.47\end{array}$&0.78$\pm$0.23&Accepted\\
\hline
\end{tabular}
\end{scriptsize}

Rem: HIP 112821 is also sometimes classified as a dwarf barium star
 
\end{table*}

\addtocounter{table}{-1}
\begin{table*}[htb]
\caption{(Continued.)}
\renewcommand{\arraystretch}{0.85}
\renewcommand{\tabcolsep}{4pt}
\begin{scriptsize}
\begin{tabular}{llllrlllcclllll}\hline
HIP&HD/DM&$\varpi_{\rm HIP}$&$N_{IAD}$&$\beta$&$P$&$e$&\multicolumn{1}{c}{$T$}
&Sol$_{\rm
HIP}$&Transit?&$\frac{\chi^2}{(N_{\rm
IAD}-9)}$&$\alpha$&\multicolumn{2}{c}{$a_0/\varpi$}&Rem\\
\cline{13-14}&&(mas)&&(\degr)&(d)&\multicolumn{3}{c}{(JD-
2\ts400\ts000)}&&&(\%)&astrom.&expected&\\\hline
\multicolumn{2}{l}{\bf Strong Ba stars}
\medskip\\
4347&5424&0.22$\pm$1.42&52&-30.9&1881.5&0.226&46202.8&5&y&1.29&51&3.7e+09$\begin{array}{l}+\infty\\0\end{array}$&1.12$\pm$0.40& \\
13055&16458&6.54$\pm$0.57&80&60.2&2018.0&0.099&46344&7&&1.20&
0&1.6$\begin{array}{l}2\\1.2\end{array}$&1.17$\pm$0.42&Accepted\\
15264&20394&2.16$\pm$1.14&48&-15.3&2226.0&0.2&47929&5&&1.39&77&8$\begin{array}{l}
46\\0\end{array}$&1.25$\pm$0.44& \\
17402&24035&3.72$\pm$0.8&70&-76.9&377.8&0.02&48842.65&5&&1.12&49&3.3e+09$\begin{array}{
l}+\infty\\0\end{array}$&0.38$\pm$0.14&*\\
23168&31487&4.54$\pm$1.21&40&29.1&1066.4&0.045&45173&5&&0.45&86&0.47$\begin{array}{l
}1.7\\0\end{array}$&0.77$\pm$0.27& \\
25452&36598&3.32$\pm$0.68&72&-85.3&2652.8&0.084&45838.95&5&&0.99&
5&4.3$\begin{array}{l}8.7\\1.3\end{array}$&1.41$\pm$0.50& \\
29099&42537&-1.13$\pm$0.93&64&-75.9&3216.2&0.156&46147.32&5&y&1.30&82&1.3e+15$\begin{array}{l}+\infty\\0\end{array}$&1.60$\pm$0.57& \\
29740&43389&-1.25$\pm$1&52&-25.8&1689.0&0.082&47222.46&5&&0.53&
0&33$\begin{array}{l}2400\\0.4\end{array}$&1.04$\pm$0.37&*\\
30338&44896&1.56$\pm$0.7&90&-56.9&628.9&0.025&48464.3&5&&0.88&
0&1.8$\begin{array}{l}3.5\\0.92\end{array}$&0.54$\pm$0.19& \\
31205&46407&8.25$\pm$0.92&80&-34.3&457.4&0.013&47677.45&O&y&0.77&
0&0.71$\begin{array}{l}1\\0.46\end{array}$&0.44$\pm$0.15&Accepted\\
32713&49641&0.73$\pm$0.88&48&-19.2&1768.0&0&46306&5&&0.63&
0&7.5$\begin{array}{l}22\\2.3\end{array}$&1.08$\pm$0.38& \\
32960&50082&4.71$\pm$0.99&40&-16.2&2896.0&0.188&45953.12&5&&1.39&44&3.9$\begin{array}{l
}8.8\\0.62\end{array}$&1.49$\pm$0.53&\\
36643&60197&1.67$\pm$0.84&92&-50.5&3243.8&0.34&46015.97&5&&0.64&59&21$\begin{array}{l}6
9\\0.66\end{array}$&1.61$\pm$0.57& \\
50006&88562&3.13$\pm$1.17&50&-25.1&1445.1&0.204&45781.71&5&&0.95&
4&3.2$\begin{array}{l}6.5\\1.3\end{array}$&0.94$\pm$0.33& \\
52271&92626&3.4$\pm$0.71&104&-50.5&918.2&0&49147.83&5&&1.05&
9&0.59$\begin{array}{l}0.99\\0.31\end{array}$&0.69$\pm$0.25&Accepted\\
56404&100503&0.67$\pm$1.2&64&-30.8&554.4&0.061&46144.83&5&&0.63&26&2.3$\begin{array}{l}
21\\0.15\end{array}$&0.50$\pm$0.18& \\
56731&101013&7.07$\pm$0.68&74&43.2&1710.9&0.195&43934&O&&1.19&
0&1.1$\begin{array}{l}1.4\\0.88\end{array}$&1.05$\pm$0.37&Accepted\\
60292&107541&5.78$\pm$1.36&68&-29.5&3569.9&0.104&44388.16&5&&0.96&
2&27$\begin{array}{l}51\\12\end{array}$&1.72$\pm$0.61&\\
68023&121447&2.21$\pm$1.02&52&-6.0&185.7&0&46922.35&5&y&1.14&19&3.5$\begin{array}{l}7.6
\\1.3\end{array}$&0.24$\pm$0.08& \\
69290&123949&.97$\pm$1.32&34&-5.6&9200.0&0.972&49144.96&5&&1.50&92&470$\begin{array}{l}
9000\\0\end{array}$&3.23$\pm$1.14& \\
94103&178717&2.9$\pm$0.95&52&32.5&2866.0&0.434&44258&5&y&0.89&71&16$\begin{array}{l}
63\\0\end{array}$&1.48$\pm$0.53& \\
101887&196445&1.49$\pm$1.54&50&-21.2&3221.3&0.237&46037.95&5&y&1.06&63&9.3$\begin{array}{l}100\\0\end{array}$&1.60$\pm$0.57& \\
103546&199939&3.16$\pm$0.75&72&57.6&584.9&0.284&45255.1&5&&0.79&26&0.44$\begin{array}{
l}0.88\\0.15\end{array}$&0.51$\pm$0.18&Accepted\\
104542&201657&4.49$\pm$1.07&64&31.6&1710.4&0.171&46154.95&5&&1.06&73&0.81$\begin{array}
{l}2.3\\0\end{array}$&1.05$\pm$0.37& \\
104684&201824&0.56$\pm$1.56&50&7.4&2837.0&0.342&47413&5&&1.79&87&2.1e+03
$\begin{array}{l}+\infty\\0\end{array}$&1.47$\pm$0.52& \\
110108&211594&4.59$\pm$1.18&26&4.4&1018.9&0.058&48538.19&5&&1.11&17&0.99$\begin{array}{
l}1.8\\0.43\end{array}$&0.74$\pm$0.26&Accepted
\medskip\\
\multicolumn{2}{l}{\bf Tc-poor S stars}
\medskip\\
5772&7351&3.21$\pm$0.82&28&19.1&4592.7&0.17&44696&5&y&0.92&88&6.9$\begin{array}{l}57
\\0\end{array}$&1.97$\pm$0.67& \\
8876&$+21^\circ255$
&-1.92$\pm$1.5&38&9.5&4137.2&0.209&43578.31&5&y&1.13&87&2.4e+13$\begin{array}{l}+\infty
\\0\end{array}$&1.84$\pm$0.66& \\
17296&22649&6.27$\pm$0.63&78&42.2&596.2&0.088&42794.5&O&y&1.22&
1&0.37$\begin{array}{l}0.52\\0.24\end{array}$&0.51$\pm$0.17&Accepted\\
25092&35155&1.32$\pm$0.99&52&-31.7&640.6&0.071&48092.41&5&y&0.79&51&0.94$\begin{array}{
l}2.7\\0.18\end{array}$&0.53$\pm$0.18&Accepted\\
32627&49368&1.65$\pm$1.11&60&-17.4&2995.9&0.357&45145.37&5&y&1.03&57&16$\begin{array}{l
}110\\0.91\end{array}$&1.48$\pm$0.50& \\
38217&63733&0$\pm$0.99&84&-39.2&1160.7&0.231&45990.92&5&&0.95&58&1.7e+14$\begin{array}{
l}+\infty\\0\end{array}$&0.79$\pm$0.27& \\
90723&170970&1.83$\pm$0.67&86&59.4&4392.0&0.084&48213.21&5&y&0.98&39&27$\begin{array}{l
}59\\8.8\end{array}$&1.91$\pm$0.66& \\
99124&191226&0.39$\pm$0.71&84&54.9&1210.4&0.19&49691.78&5&&0.98&23&11$\begin{array}{l}6
20\\0.14\end{array}$&0.81$\pm$0.27& \\
99312&191589&2.25$\pm$0.77&84&52.0&377.3&0.253&48844.02&5&&0.80&82&0.58$\begin{array}{l
}2.4\\0\end{array}$&0.37$\pm$0.13&Accepted\\
115965&$+28^\circ4592$
&1.72$\pm$1.26&64&29.6&1252.9&0.091&48161.32&5&&0.78&72&1.4$\begin{array}{l}8.9\\0\end{array}$&0.83$\pm$0.28& \\
\hline
\end{tabular}
\end{scriptsize}

HIP 29740: $\alpha=0$ but the solutions are totally unrealistic\\
HIP 17402: $\varpi$ not accurately determined, 
since orbital period is close to 1~yr.

\end{table*}

The following columns characterize the quality of the solution obtained from 
the minimization process.  Column~11 provides $\chi^2/(N_{\rm IAD}-9)$,  
which should be of the order of unity if the internal error $\sigma$
(Eq.~\ref{Eq:sigma}) on the abscissae has been correctly evaluated by the 
reduction consortia and if the model provides an adequate representation of 
the data.  The first risk error associated with rejecting the 
null hypothesis that the orbital and Hipparcos 5-parameter solutions are 
identical (Eq.~\ref{Eq:F}) is given in column~12. Low values of $\alpha$ are 
generally associated with Hipparcos solutions of the G, X or O types (as 
listed in column~9; see the caption to Table~\ref{Tab:criteria} for more 
details), since the orbital motion is then large enough to have been noticed
already by NDAC or FAST.  Columns~13 and 14 compare the 
$a_0/\varpi$ ratio derived from the orbital solution to its expected 
value from Eq.~\ref{Eq:a0varpi}.  In columns~11 to 13,  the data refer 
to the orbital solution obtained by combining NDAC and FAST data. 
For dwarf barium stars, the masses used to estimate $a_0/\varpi$ 
according to Eq.~\ref{Eq:a0varpi} are $M_2 = 0.62\pm 0.04$~\Msun, 
whereas $M_1$ is derived from the spectroscopic 
gravity, with an estimated error of 0.05\Msun\ (North, priv. comm.). 
According to the statistical analysis of the spectroscopic mass 
functions performed by \cite*{Jorissen-1998}, $M_1$ and $M_2$ pairs (expressed
in \Msun) of ($1.7\pm 0.2,  0.62\pm 0.04$), ($2.1\pm 0.2,  0.62\pm 0.04$) and 
($1.8\pm 0.2,  0.62\pm 0.04$) have been 
adopted for strong barium stars, mild  barium stars and Tc-poor S 
stars, respectively. The same analysis performed by 
\cite*{McClure-1990} for CH stars yielded  $M_1 = 1.0\pm 0.1$~\Msun\ and 
$M_2 =  0.62\pm 0.04$~\Msun.  

Astrometric orbits were accepted when $\alpha$ is smaller than 10 per cent 
and the expected $a_0/\varpi$ value falls within the $1\sigma$ confidence 
interval.  A few cases not fulfilling these criteria were nevertheless 
accepted after visual inspection of the orbital arc.

Examination of Table~\ref{Tab:criteria} reveals that the following criteria 
need to be fulfilled in order 
to be able to extract a reliable astrometric orbit from the 
Hipparcos data: $\varpi \ga 5$~mas, $P \la 10$~yr, $N  \ga 25$. 
The success rate is as follows for the various samples: dwarf barium stars 
(9/13), mild barium stars (3/24), strong barium stars (6/26), Tc-poor S 
stars (3/10) and CH stars (2/8). The high success rate for dwarf barium 
stars naturally results from the fact that these dwarf stars are on average 
closer from the sun than the giant stars.

\setlength{\tabcolsep}{4pt}
\begin{sidewaystable*}
\caption{\label{Tab:orbits}
Astrometric orbital elements for various classes of CPRS. The
first line provides the data from the HIPPARCOS catalogue or from
the spectroscopic orbit (label H/S in column 2). The following lines 
provide the orbital solutions derived in the present work from the FAST, 
NDAC and combined FAST+NDAC IAD (labels F, N and A, respectively in column 2).
When available, the orbital solution from the transit data (label T in 
column 2) or from the DMSA/O (label O) are given next.  
The columns $\varpi$, $a_0/\varpi$ and $K_1$
contain the nominal value followed by the boundaries of the `$1\sigma$' 
confidence interval (computed by propagating the 1$\sigma$ errors on $a_0$,
$\varpi'$ and $i$ as if they were uncorrelated). 
An asterisk after the HIP number refers to a note 
at the end of the Table}
\renewcommand{\arraystretch}{0.6}
\begin{tabular}{lccccccccccccccc}\hline
HIP&Sol&$\varpi$&$\mu_{\alpha}^*$&$\mu_{\delta}$&$a_0$&$i$&$\omega$&$\Omega$&$a
_0/\varpi$&$K_1$&N&$\chi^2$&$\chi^2$&$\alpha$\\
&&(mas)&(mas/y)&(mas/y)&(mas)&(\degr)&(\degr)&(\degr)&(AU)&(km/s)&&HIP&Orb&(\%)
\\\hline
\medskip\\
\multicolumn{4}{l}{\bf CH stars}
\medskip\\
53763*& H/S &5.29$\pm$1.47& -24.16$\pm$1.18& -15.03$\pm$1.2& & & -- & &
0.44$\pm$0.22& 5.42$\pm$0.28&67&&&\\
&F& 2.88$\begin{array}{l}8.51\\0.97\end{array}$&-23.28$\pm$1.3& -14.66$\pm$1.2&
4.92$\pm$1.9& 62.6$\pm$22.9& 113.0$\pm$44.3&
247.6$\pm$44.1&1.71$\begin{array}{l}6.98\\0.36\end{array}$&50.25$\begin{array}{
l}230.64\\7.62\end{array}$ &32& 33.7&22.8&  0\\
&N& 1.59$\begin{array}{l}15.37\\0.16\end{array}$&-24.22$\pm$1.6& -15.43$\pm$1.5&
5.21$\pm$2.5& 79.4$\pm$19.7& 100.2$\pm$34.7&
258.4$\pm$32.6&3.27$\begin{array}{l}47.23\\0.18\end{array}$&106.60$\begin{array}{l}1565.08\\5.08\end{array}$ &33& 28.9&21.0&  0\\
&A& 2.70$\begin{array}{l}8.18\\0.89\end{array}$&-23.61$\pm$1.2& -15.13$\pm$1.2&
4.91$\pm$1.9& 68.7$\pm$19.3& 103.5$\pm$37.6&
254.2$\pm$36.0&1.81$\begin{array}{l}7.67\\0.36\end{array}$&56.00$\begin{array}{l}254.06\\9.16\end{array}$ &70& 61.2&51.4&  1
\medskip\\
104486& H/S &4.93$\pm$0.84& -43$\pm$0.66& 29.96$\pm$0.52& & & 133$\pm$20& &
1.2$\pm$0.58& 5.3$\pm$0.19&99&&&\\
&F& 4.64$\begin{array}{l}5.63\\3.83\end{array}$&-48.91$\pm$2.9& 32.21$\pm$2.1&
8.18$\pm$1.4& 71.3$\pm$19.0& 74.7$\pm$22.2&
103.1$\pm$10.6&1.76$\begin{array}{l}2.49\\1.21\end{array}$&12.46$\begin{array}{
l}18.62\\7.15\end{array}$ &48& 90.2&69.2&  0\\
&N& 6.56$\begin{array}{l}7.67\\5.60\end{array}$&-48.78$\pm$3.2& 32.23$\pm$2.4&
7.34$\pm$1.5& 66.7$\pm$24.9& 86.5$\pm$30.6&
97.7$\pm$16.3&1.12$\begin{array}{l}1.58\\0.76\end{array}$&7.67$\begin{array}{l}
11.76\\3.80\end{array}$ &49& 68.4&49.1&  0\\
&A& 5.09$\begin{array}{l}6.01\\4.32\end{array}$&-48.28$\pm$2.7& 32.43$\pm$2.0&
7.46$\pm$1.2& 64.8$\pm$21.6& 75.7$\pm$24.5&
104.4$\pm$13.6&1.46$\begin{array}{l}2.01\\1.04\end{array}$&9.89$\begin{array}{l
}14.98\\5.30\end{array}$ &100& 145.0&123.3&  0
\medskip\\
\multicolumn{4}{l}{\bf Dwarf Ba stars}
\medskip\\
32894& H/S &14.11$\pm$1.96& 32.62$\pm$1.36& 106.64$\pm$1.86& & & 223$\pm$12& &
0.89$\pm$0.46& 9.7$\pm$0.2&77&&&\\
&F& 14.17$\begin{array}{l}15.41\\13.04\end{array}$&34.81$\pm$1.0& 108.50$\pm$1.4&
12.29$\pm$1.2& 108.4$\pm$6.5& 226.8$\pm$6.5&
69.7$\pm$8.7&0.87$\begin{array}{l}1.03\\0.72\end{array}$&9.85$\begin{array}{l}1
2.11\\7.83\end{array}$ &38& 159.4&37.6&  0\\
&N& 13.57$\begin{array}{l}15.03\\12.24\end{array}$&35.16$\pm$1.2& 108.49$\pm$1.8&
11.43$\pm$1.4& 111.7$\pm$8.7& 219.8$\pm$8.2&
64.1$\pm$12.1&0.84$\begin{array}{l}1.05\\0.67\end{array}$&9.38$\begin{array}{l}
12.27\\6.87\end{array}$ &39& 113.8&40.8&  0\\
&A& 14.25$\begin{array}{l}15.45\\13.15\end{array}$&35.07$\pm$0.9& 108.41$\pm$1.4&
11.99$\pm$1.2& 109.0$\pm$6.5& 225.1$\pm$6.2&
69.2$\pm$8.9&0.84$\begin{array}{l}1.00\\0.70\end{array}$&9.53$\begin{array}{l}1
1.70\\7.58\end{array}$ &78& 191.3&72.2&  0
\medskip\\
49166& H/S &7.9$\pm$1.39& -65.45$\pm$0.85& 70.1$\pm$1.1& & & 128$\pm$3& &
0.39$\pm$0.19& 12.28$\pm$0.1&87&&&\\
&F& 6.94$\begin{array}{l}8.62\\5.59\end{array}$&-66.42$\pm$1.5& 70.59$\pm$1.2&
3.81$\pm$1.6& 113.6$\pm$33.1& 107.7$\pm$30.7&
341.6$\pm$23.2&0.55$\begin{array}{l}0.97\\0.26\end{array}$&20.36$\begin{array}{
l}39.08\\5.73\end{array}$ &42& 46.5&40.5&  3\\
&N& 7.40$\begin{array}{l}9.27\\5.91\end{array}$&-66.24$\pm$1.6& 70.45$\pm$1.5&
4.42$\pm$1.8& 88.8$\pm$28.1& 87.8$\pm$23.9&
356.7$\pm$15.1&0.60$\begin{array}{l}1.04\\0.29\end{array}$&24.15$\begin{array}{
l}42.24\\10.16\end{array}$ &43& 48.8&42.2&  2\\
&A& 7.26$\begin{array}{l}8.83\\5.97\end{array}$&-66.38$\pm$1.4& 70.50$\pm$1.2&
3.44$\pm$1.5& 109.1$\pm$32.3& 99.7$\pm$30.6&
338.1$\pm$21.3&0.47$\begin{array}{l}0.83\\0.22\end{array}$&18.11$\begin{array}{
l}33.52\\5.54\end{array}$ &90& 88.2&83.4& 34\\
\hline
\end{tabular}
\end{sidewaystable*}

\addtocounter{table}{-1}
\begin{sidewaystable*}
\caption{(Continued.)}
\renewcommand{\arraystretch}{0.6}
\begin{tabular}{lccccccccccccccc}\hline
HIP&Sol&$\varpi$&$\mu_{\alpha}^*$&$\mu_{\delta}$&$a_0$&$i$&$\omega$&$\Omega$&$a
_0/\varpi$&$K_1$&N&$\chi^2$&$\chi^2$&$\alpha$\\
&&(mas)&(mas/y)&(mas/y)&(mas)&(\degr)&(\degr)&(\degr)&(AU)&(km/s)&&HIP&Orb&(\%)
\\\hline
\medskip\\
\multicolumn{4}{l}{\bf Dwarf Ba stars (cont.)}
\medskip\\
50805& H/S &23.42$\pm$0.93& -26.77$\pm$0.67& 55.41$\pm$1.36& & & 137$\pm$4& &
0.67$\pm$0.3& 9.67$\pm$0.1&58&&&\\
&F& 23.43$\begin{array}{l}24.36\\22.52\end{array}$&-27.78$\pm$0.8& 56.42$\pm$0.9&
15.75$\pm$1.0& 102.1$\pm$3.0& 144.7$\pm$4.5&
343.1$\pm$4.0&0.67$\begin{array}{l}0.74\\0.61\end{array}$&10.79$\begin{array}{l
}12.04\\9.60\end{array}$ &29& 296.4&22.2&  0\\
&N& 22.94$\begin{array}{l}24.01\\21.91\end{array}$&-25.91$\pm$0.8& 56.12$\pm$1.1&
15.28$\pm$1.2& 99.8$\pm$3.3& 147.7$\pm$5.5&
343.2$\pm$4.4&0.67$\begin{array}{l}0.75\\0.59\end{array}$&10.77$\begin{array}{l
}12.23\\9.39\end{array}$ &29& 202.9&16.3&  0\\
&A& 23.40$\begin{array}{l}24.28\\22.54\end{array}$&-26.89$\pm$0.7& 56.51$\pm$0.9&
15.59$\pm$0.9& 100.6$\pm$2.8& 146.3$\pm$4.3&
343.8$\pm$3.7&0.67$\begin{array}{l}0.73\\0.60\end{array}$&10.75$\begin{array}{l
}11.92\\9.63\end{array}$ &60& 330.2&42.0&  0\\
&O&&&&15.76$\pm$1.0& 100.2$\pm$2.9& -- & 344.1$\pm$3.6& 0.67$\pm$0.05\\
&T& 23.5$\begin{array}{l}24.6\\22.3\end{array}$&-26.3$\pm$0.83& 56.4$\pm$0.98&
16$\pm$1.3& 99$\pm$3.6& 142$\pm$5.3&
342$\pm$4.6&0.69$\begin{array}{l}0.78\\0.61\end{array}$&&98&&&
\medskip\\
60299& H/S &5.02$\pm$1.06& -147.79$\pm$0.96& 9.32$\pm$0.7& & & 69$\pm$48& &
0.83$\pm$0.26& 2.32$\pm$0.16&52&&&\\
&F& 5.37$\begin{array}{l}6.59\\4.37\end{array}$&-141.76$\pm$2.1& 10.25$\pm$1.9&
9.35$\pm$2.5& 116.9$\pm$13.1& 127.5$\pm$13.3&
266.6$\pm$18.9&1.74$\begin{array}{l}2.72\\1.04\end{array}$&12.56$\begin{array}{
l}21.33\\6.41\end{array}$ &25& 62.4&15.3&  0\\
&N& 6.02$\begin{array}{l}7.38\\4.91\end{array}$&-139.47$\pm$2.6& 11.74$\pm$1.5&
12.02$\pm$3.0& 118.0$\pm$10.0& 113.9$\pm$13.0&
258.5$\pm$15.0&1.99$\begin{array}{l}3.05\\1.22\end{array}$&14.24$\begin{array}{
l}23.47\\7.80\end{array}$ &25& 57.6&15.9&  0\\
&A& 5.64$\begin{array}{l}6.77\\4.69\end{array}$&-141.41$\pm$1.9& 11.20$\pm$1.2&
9.95$\pm$2.4& 123.3$\pm$11.6& 124.1$\pm$14.3&
264.1$\pm$16.6&1.76$\begin{array}{l}2.63\\1.12\end{array}$&11.93$\begin{array}{
l}19.74\\6.40\end{array}$ &50& 96.3&38.1&  0
\medskip\\
62409*& H/S &8.2$\pm$1.28& -136.5$\pm$1.63& -19.28$\pm$0.79& & & 186$\pm$24& &
1.3$\pm$0.65& 7.17$\pm$0.22&57&&&\\
&[F]& 7.66$\begin{array}{l}9.31\\6.30\end{array}$&-137.04$\pm$7.9& -15.07$\pm$3.7&
5.94$\pm$4.1& 81.4$\pm$42.6& 167.3$\pm$25.3&
182.8$\pm$80.2&0.78$\begin{array}{l}1.60\\0.19\end{array}$&4.69$\begin{array}{l
}9.80\\0.74\end{array}$ &28& 21.9&15.6&  0\\
&N& 8.17$\begin{array}{l}10.18\\6.56\end{array}$&-126.05$\pm$8.9& -16.19$\pm$3.8&
11.88$\pm$8.8& 120.8$\pm$32.5& 276.0$\pm$35.5&
293.1$\pm$31.2&1.45$\begin{array}{l}3.15\\0.30\end{array}$&7.64$\begin{array}{l
}19.26\\0.84\end{array}$ &29& 36.3&17.5&  0\\
&[A]& 7.68$\begin{array}{l}9.25\\6.38\end{array}$&-136.08$\pm$7.2& -15.09$\pm$3.3&
6.71$\pm$4.7& 94.2$\pm$34.1& 168.2$\pm$21.0&
165.1$\pm$57.7&0.87$\begin{array}{l}1.79\\0.22\end{array}$&5.33$\begin{array}{l
}10.95\\1.04\end{array}$ &58& 70.2&63.0& 25
\medskip\\
69176& H/S &8.75$\pm$1.39& 34.08$\pm$1.22& -29.9$\pm$1.12& & & 192$\pm$18& &
0.44$\pm$0.16& 12$\pm$0.43&55&&&\\
&F& 10.36$\begin{array}{l}12.46\\8.61\end{array}$&35.36$\pm$1.6& -28.80$\pm$1.2&
4.64$\pm$1.1& 52.6$\pm$31.0& 203.8$\pm$32.2&
17.0$\pm$40.9&0.45$\begin{array}{l}0.67\\0.28\end{array}$&8.48$\begin{array}{l}
15.83\\2.48\end{array}$ &25& 46.1&23.6&  0\\
&N& 9.80$\begin{array}{l}12.31\\7.79\end{array}$&33.08$\pm$1.8& -30.76$\pm$1.4&
5.07$\pm$1.4& 63.8$\pm$27.6& 191.5$\pm$22.9&
30.9$\pm$30.2&0.52$\begin{array}{l}0.84\\0.29\end{array}$&11.05$\begin{array}{l
}19.92\\4.13\end{array}$ &27& 37.3&15.6&  0\\
&A& 10.16$\begin{array}{l}12.19\\8.46\end{array}$&34.38$\pm$1.5& -29.26$\pm$1.1&
4.81$\pm$1.1& 53.9$\pm$28.4& 203.5$\pm$28.2&
8.5$\pm$35.5&0.47$\begin{array}{l}0.69\\0.31\end{array}$&9.11$\begin{array}{l}1
6.37\\3.15\end{array}$ &54& 82.3&55.9&  0\\
\hline
\end{tabular}
\end{sidewaystable*}

\addtocounter{table}{-1}
\begin{sidewaystable*}
\caption{(Continued.)}
\renewcommand{\arraystretch}{0.6}
\begin{tabular}{lccccccccccccccc}\hline
HIP&Sol&$\varpi$&$\mu_{\alpha}^*$&$\mu_{\delta}$&$a_0$&$i$&$\omega$&$\Omega$&$a
_0/\varpi$&$K_1$&N&$\chi^2$&$\chi^2$&$\alpha$\\

&&(mas)&(mas/y)&(mas/y)&(mas)&(\degr)&(\degr)&(\degr)&(AU)&(km/s)&&HIP&Orb&(\%)
\\\hline
\medskip\\
\multicolumn{4}{l}{\bf Dwarf Ba stars (cont.)}
\medskip\\
105969& H/S &16.61$\pm$1.78& 200.64$\pm$1.56& -93.14$\pm$1.52& & & 192$\pm$21& &
0.77$\pm$0.33& 3.29$\pm$0.15&77&&&\\
&F& 16.10$\begin{array}{l}16.96\\15.29\end{array}$&200.75$\pm$0.8& -93.48$\pm$0.7&
12.27$\pm$0.9& 132.7$\pm$6.9& 225.9$\pm$11.2&
273.6$\pm$10.9&0.76$\begin{array}{l}0.86\\0.67\end{array}$&7.00$\begin{array}{l
}8.75\\5.42\end{array}$ &38& 334.0&24.7&  0\\
&N& 15.92$\begin{array}{l}16.84\\15.05\end{array}$&200.27$\pm$0.7& -92.23$\pm$0.8&
12.07$\pm$1.0& 132.1$\pm$7.5& 239.6$\pm$12.0&
282.7$\pm$11.8&0.76$\begin{array}{l}0.87\\0.66\end{array}$&7.03$\begin{array}{l
}8.90\\5.34\end{array}$ &39& 290.9&40.2&  0\\
&A& 16.22$\begin{array}{l}17.01\\15.46\end{array}$&200.25$\pm$0.6& -93.07$\pm$0.6&
12.12$\pm$0.8& 133.1$\pm$6.6& 233.0$\pm$10.8&
277.5$\pm$10.6&0.75$\begin{array}{l}0.84\\0.66\end{array}$&6.81$\begin{array}{l
}8.41\\5.36\end{array}$ &80& 418.7&72.1&  0
\medskip\\
107818& H/S &7.53$\pm$1.51& 16.16$\pm$1.24& -36.55$\pm$0.8& & & 281$\pm$42& &
0.58$\pm$0.22& 10.4$\pm$0.19&50&&&\\
&F& 8.66$\begin{array}{l}10.54\\7.12\end{array}$&16.50$\pm$1.6& -36.90$\pm$1.0&
3.09$\pm$1.4& 107.0$\pm$40.5& 306.0$\pm$27.3&
153.4$\pm$43.2&0.36$\begin{array}{l}0.63\\0.16\end{array}$&5.54$\begin{array}{l
}10.22\\1.41\end{array}$ &25& 22.5&16.3&  1\\
&N& 8.64$\begin{array}{l}10.79\\6.92\end{array}$&16.89$\pm$1.8& -36.43$\pm$1.1&
4.23$\pm$2.3& 126.7$\pm$31.0& 340.0$\pm$55.9&
226.8$\pm$48.4&0.49$\begin{array}{l}0.95\\0.18\end{array}$&6.37$\begin{array}{l
}15.29\\1.09\end{array}$ &25& 24.8&15.4&  0\\
&A& 8.80$\begin{array}{l}10.61\\7.30\end{array}$&16.95$\pm$1.5& -36.35$\pm$1.0&
3.44$\pm$1.4& 107.3$\pm$32.4& 302.1$\pm$28.5&
175.7$\pm$37.3&0.39$\begin{array}{l}0.66\\0.19\end{array}$&6.06$\begin{array}{l
}10.71\\2.05\end{array}$ &52& 42.7&35.1&  7
\medskip\\
116233*& H/S &8.83$\pm$1.21& 27.08$\pm$1.69& -22.19$\pm$1.14& & & 196$\pm$17& &
0.97$\pm$0.37& 7.42$\pm$0.26&54&&&\\
&F& 9.76$\begin{array}{l}11.29\\8.43\end{array}$&26.61$\pm$4.0& -9.47$\pm$3.3&
11.10$\pm$1.4& 26.7$\pm$50.4& 355.1$\pm$91.1&
158.1$\pm$87.7&1.14$\begin{array}{l}1.48\\0.86\end{array}$&3.98$\begin{array}{l
}11.24\\0.00\end{array}$ &23& 149.7&4.6&  0\\
&N& 10.28$\begin{array}{l}12.34\\8.56\end{array}$&44.68$\pm$17.8& -16.68$\pm$6.3&
14.91$\pm$9.0& 76.8$\pm$17.4& 312.1$\pm$36.2&
249.5$\pm$32.7&1.45$\begin{array}{l}2.79\\0.48\end{array}$&11.00$\begin{array}{
l}21.75\\3.22\end{array}$ &27& 113.4&17.6&  0\\
&A& 9.57$\begin{array}{l}10.97\\8.35\end{array}$&26.58$\pm$3.8& -9.98$\pm$3.2&
11.11$\pm$1.4& 23.0$\pm$54.2& 201.9$\pm$118.9&
312.0$\pm$115.1&1.16$\begin{array}{l}1.50\\0.89\end{array}$&3.53$\begin{array}{
l}11.38\\0.00\end{array}$ &56& 209.5&39.3&  0
\medskip\\
118266*& H/S &6.95$\pm$1.45& -30.31$\pm$1.37& -121.84$\pm$0.86& & & 250$\pm$44&
& 0.4$\pm$0.19& 8.55$\pm$0.39&79&&&\\
&F& 8.56$\begin{array}{l}10.90\\6.72\end{array}$&-29.70$\pm$1.5& -120.64$\pm$1.1&
2.60$\pm$1.7& 45.2$\pm$66.7& 360.6$\pm$73.3&
219.1$\pm$68.2&0.30$\begin{array}{l}0.64\\0.09\end{array}$&7.63$\begin{array}{l
}22.50\\0.00\end{array}$ &39& 42.2&37.3&  7\\
&N& 7.45$\begin{array}{l}10.35\\5.35\end{array}$&-30.71$\pm$1.7& -121.93$\pm$1.3&
3.11$\pm$2.1& 112.1$\pm$29.6& 267.6$\pm$48.7&
57.0$\pm$38.7&0.42$\begin{array}{l}0.98\\0.10\end{array}$&13.71$\begin{array}{l
}34.53\\2.12\end{array}$ &39& 25.9&22.6&  5\\
&A& 8.56$\begin{array}{l}10.84\\6.76\end{array}$&-30.08$\pm$1.4& -120.97$\pm$1.0&
2.30$\pm$1.4& 60.1$\pm$55.6& 205.2$\pm$47.3&
18.5$\pm$53.7&0.27$\begin{array}{l}0.55\\0.08\end{array}$&8.24$\begin{array}{l}
19.46\\0.23\end{array}$ &80& 74.7&71.0& 45\\
\hline
\end{tabular}

Remarks:\\
HIP 62409: Only NDAC yields a realistic orbit.\\
HIP 116233, HIP 118266: Orbits not well defined.\\
HIP 53763: orbit not well defined with large uncertainty on $\varpi$
since $P$ is close to 1 yr.
\end{sidewaystable*}

\addtocounter{table}{-1}
\begin{sidewaystable*}
\caption{(Continued.)}
\renewcommand{\arraystretch}{0.6}
\begin{tabular}{lccccccccccccccc}\hline
HIP&Sol&$\varpi$&$\mu_{\alpha}^*$&$\mu_{\delta}$&$a_0$&$i$&$\omega$&$\Omega$&$a
_0/\varpi$&$K_1$&N&$\chi^2$&$\chi^2$&$\alpha$\\
&&(mas)&(mas/y)&(mas/y)&(mas)&(\degr)&(\degr)&(\degr)&(AU)&(km/s)&&HIP&Orb&(\%)
\\\hline
\medskip\\
\multicolumn{4}{l}{\bf Mild Ba stars}
\medskip\\
36042*& H/S &2.36$\pm$0.97& 2.74$\pm$0.94& -0.74$\pm$0.61& & & 17$\pm$4.6& &
0.5$\pm$0.15& 6.91$\pm$0.14&56&&&\\
&F& 2.78$\begin{array}{l}4.11\\1.89\end{array}$&2.49$\pm$1.0& -0.51$\pm$0.7&
5.79$\pm$1.2& 94.0$\pm$9.7& 81.8$\pm$13.3&
289.1$\pm$8.1&2.08$\begin{array}{l}3.71\\1.11\end{array}$&34.39$\begin{array}{l
}61.53\\17.95\end{array}$ &28& 33.7&15.8&  0\\
&N& 2.00$\begin{array}{l}3.56\\1.13\end{array}$&2.78$\pm$1.1& -0.20$\pm$0.7&
3.57$\pm$1.3& 119.1$\pm$22.7& 61.4$\pm$30.5&
296.7$\pm$23.4&1.78$\begin{array}{l}4.32\\0.64\end{array}$&25.86$\begin{array}{
l}71.18\\6.59\end{array}$ &28& 28.0&18.5&  0\\
&A& 2.44$\begin{array}{l}3.69\\1.62\end{array}$&2.60$\pm$1.0& -0.31$\pm$0.6&
4.86$\pm$1.1& 101.0$\pm$11.5& 75.2$\pm$14.9&
295.8$\pm$9.9&1.99$\begin{array}{l}3.69\\1.02\end{array}$&32.36$\begin{array}{l
}61.16\\15.56\end{array}$ &56& 63.1&46.9&  1
\medskip\\
105881*& H/S &8.19$\pm$0.9& -2.61$\pm$0.96& 18.88$\pm$0.42& & & 239.51$\pm$10& &
1.2$\pm$0.35& 2.763$\pm$0.161&51&&&\\
&F& 9.05$\begin{array}{l}10.23\\8.00\end{array}$&-0.55$\pm$6.6& 26.59$\pm$3.4&
9.77$\pm$4.7& 102.8$\pm$17.9& 228.6$\pm$12.5&
187.6$\pm$45.2&1.08$\begin{array}{l}1.81\\0.50\end{array}$&5.02$\begin{array}{l
}8.62\\2.04\end{array}$ &25& 29.8&18.9&  0\\
&N& 7.71$\begin{array}{l}9.08\\6.54\end{array}$&4.71$\pm$7.4& 23.37$\pm$3.8&
9.24$\pm$8.3& 119.8$\pm$31.2& 269.4$\pm$32.3&
237.5$\pm$45.1&1.20$\begin{array}{l}2.68\\0.11\end{array}$&4.96$\begin{array}{l
}12.77\\0.24\end{array}$ &26& 24.2&12.8&  0\\
&A& 8.60$\begin{array}{l}9.67\\7.65\end{array}$&-0.57$\pm$5.8& 25.30$\pm$3.1&
8.56$\pm$4.3& 111.7$\pm$20.7& 233.8$\pm$19.8&
190.7$\pm$50.0&1.00$\begin{array}{l}1.69\\0.44\end{array}$&4.41$\begin{array}{l
}8.05\\1.53\end{array}$ &54& 50.2&41.4&  6\\
&T& 7.7$\begin{array}{l}7.8\\7.5\end{array}$&9.5$\pm$0.86& 24.5$\pm$0.49&
12.3$\pm$0.89& 112$\pm$2.6& 261$\pm$2.7&
240$\pm$3.5&1.6$\begin{array}{l}1.75\\1.46\end{array}$&&86&&&
\medskip\\
117607*& H/S &4.61$\pm$0.95& 2.88$\pm$1.29& 3.32$\pm$0.92& & & 157.61$\pm$25& &
0.78$\pm$0.23& 3.63$\pm$0.08&60&&&\\
&F& 4.35$\begin{array}{l}5.59\\3.39\end{array}$&6.43$\pm$3.6& 0.57$\pm$1.4&
3.55$\pm$1.7& 116.0$\pm$38.8& 139.1$\pm$56.0&
128.7$\pm$57.5&0.82$\begin{array}{l}1.56\\0.33\end{array}$&6.18$\begin{array}{l
}13.13\\1.17\end{array}$ &29& 104.1&18.5&  0\\
&N& 3.91$\begin{array}{l}5.40\\2.84\end{array}$&-0.52$\pm$3.2& -3.07$\pm$1.6&
6.83$\pm$2.6& 83.8$\pm$15.5& 138.4$\pm$21.1&
214.8$\pm$18.6&1.74$\begin{array}{l}3.31\\0.79\end{array}$&14.61$\begin{array}{
l}27.93\\6.18\end{array}$ &31& 111.5&15.0&  0\\
&A& 3.92$\begin{array}{l}5.07\\3.03\end{array}$&3.33$\pm$2.7& -0.58$\pm$1.3&
3.41$\pm$1.0& 82.3$\pm$37.1& 154.6$\pm$23.4&
173.3$\pm$42.4&0.87$\begin{array}{l}1.46\\0.47\end{array}$&7.27$\begin{array}{l
}12.30\\2.83\end{array}$ &62& 150.2&39.9&  0\\
\hline
\end{tabular}

Remark:
\begin{description}
\item{HIP 36042}: large discrepancy between astrometric and
astrophysical $a_0/\varpi$.  However, the orbit has been retained
since the correlation between the 9 parameters
is very low (the average efficiency of the
three solutions is 0.69; see Eq.~\protect\ref{Eq:efficiency}).
\item{HIP~105881}: orbit not well defined.
\item{HIP~117607}: although the orbit is not well defined, it has been 
  kept in the Table to illustrate the correlations that may appear
  between $a_0$ and $\mu$
\end{description}
\end{sidewaystable*}

\addtocounter{table}{-1}
\begin{sidewaystable*}
\caption{(Continued.)}
\renewcommand{\arraystretch}{0.6}
\begin{tabular}{lccccccccccccccc}\hline
HIP&Sol&$\varpi$&$\mu_{\alpha}^*$&$\mu_{\delta}$&$a_0$&$i$&$\omega$&$\Omega$&$a
_0/\varpi$&$K_1$&N&$\chi^2$&$\chi^2$&$\alpha$\\
&&(mas)&(mas/y)&(mas/y)&(mas)&(\degr)&(\degr)&(\degr)&(AU)&(km/s)&&HIP&Orb&(\%)
\\\hline
\medskip\\
\multicolumn{4}{l}{\bf Strong Ba stars}
\medskip\\
13055& H/S &6.54$\pm$0.57& 18.78$\pm$0.54& -78.4$\pm$0.62& & & 120$\pm$11& &
1.2$\pm$0.42& 5.83$\pm$0.13&78&&&\\
&F& 6.62$\begin{array}{l}7.28\\6.01\end{array}$&10.76$\pm$2.2& -76.39$\pm$2.9&
9.90$\pm$2.0& 86.3$\pm$13.4& 117.5$\pm$14.4&
281.1$\pm$15.0&1.50$\begin{array}{l}1.98\\1.09\end{array}$&8.09$\begin{array}{l
}10.71\\5.62\end{array}$ &38& 130.8&46.8&  0\\
&N& 6.96$\begin{array}{l}7.72\\6.27\end{array}$&8.95$\pm$2.4& -76.21$\pm$3.4&
11.95$\pm$2.1& 89.2$\pm$11.8& 106.6$\pm$12.8&
278.5$\pm$15.5&1.72$\begin{array}{l}2.24\\1.28\end{array}$&9.31$\begin{array}{l
}12.15\\6.75\end{array}$ &40& 110.1&37.6&  0\\
&A& 6.81$\begin{array}{l}7.41\\6.25\end{array}$&10.00$\pm$2.0& -76.29$\pm$2.6&
10.83$\pm$1.8& 87.1$\pm$10.8& 113.1$\pm$11.7&
280.5$\pm$12.6&1.59$\begin{array}{l}2.01\\1.22\end{array}$&8.61$\begin{array}{l
}10.91\\6.44\end{array}$ &80& 192.9&85.3&  0
\medskip\\
31205& H/S &8.25$\pm$0.92& -15.95$\pm$0.89& 6.22$\pm$0.84& & & 73.73$\pm$32& &
0.44$\pm$0.15& 9.03$\pm$0.07&79&&&\\
&F& 7.09$\begin{array}{l}8.55\\5.88\end{array}$&-16.99$\pm$1.2& 6.62$\pm$1.0&
4.26$\pm$1.1& 74.1$\pm$12.1& 98.5$\pm$20.4&
299.7$\pm$13.2&0.60$\begin{array}{l}0.91\\0.37\end{array}$&13.74$\begin{array}{
l}21.52\\7.83\end{array}$ &39& 61.8&26.8&  0\\
&N& 6.05$\begin{array}{l}7.66\\4.77\end{array}$&-18.35$\pm$1.4& 7.47$\pm$1.1&
6.04$\pm$1.2& 86.6$\pm$8.8& 120.7$\pm$14.7&
295.9$\pm$8.9&1.00$\begin{array}{l}1.52\\0.63\end{array}$&23.70$\begin{array}{l
}36.23\\14.57\end{array}$ &40& 58.6&23.9&  0\\
&A& 6.61$\begin{array}{l}7.93\\5.51\end{array}$&-17.36$\pm$1.1& 6.77$\pm$0.9&
4.72$\pm$1.0& 79.5$\pm$9.5& 109.5$\pm$16.2&
295.2$\pm$9.9&0.71$\begin{array}{l}1.04\\0.47\end{array}$&16.69$\begin{array}{l
}24.78\\10.41\end{array}$ &80& 96.3&54.7&  0\\
&O&&&&3.94$\pm$0.9&74.8$\pm$11.9&&313.8$\pm$12.7& 0.48$\pm$0.11\\
&T& 6.51$\begin{array}{l}7.2\\5.8\end{array}$&-17.6$\pm$0.58& 6.7$\pm$0.49&
5.22$\pm$0.63& 85$\pm$3.7& 118$\pm$6.7&297$\pm$4.9&
0.78$\begin{array}{l}0.98\\0.61\end{array}$&&191&&&
\medskip\\
52271& H/S &3.4$\pm$0.71& -46.75$\pm$0.49& 2.08$\pm$0.52& & & -- & &
0.69$\pm$0.25& 7.64$\pm$0.09&101&&&\\
&F& 3.76$\begin{array}{l}4.67\\3.03\end{array}$&-47.21$\pm$0.6& 1.99$\pm$0.6&
2.04$\pm$0.8& 69.4$\pm$26.1& 57.5$\pm$25.9&
30.6$\pm$24.9&0.54$\begin{array}{l}0.94\\0.26\end{array}$&6.03$\begin{array}{l}
11.19\\2.14\end{array}$ &49& 42.8&33.4&  0\\
&N& 3.33$\begin{array}{l}4.33\\2.55\end{array}$&-46.70$\pm$0.6& 2.33$\pm$0.7&
2.17$\pm$0.9& 106.7$\pm$25.8& 49.0$\pm$27.6&
38.2$\pm$24.3&0.65$\begin{array}{l}1.21\\0.29\end{array}$&7.39$\begin{array}{l}
14.32\\2.51\end{array}$ &52& 54.4&47.0&  0\\
&A& 3.43$\begin{array}{l}4.26\\2.77\end{array}$&-46.77$\pm$0.5& 1.87$\pm$0.6&
2.03$\pm$0.8& 76.1$\pm$22.9& 49.3$\pm$22.4&
35.2$\pm$20.3&0.59$\begin{array}{l}1.01\\0.30\end{array}$&6.79$\begin{array}{l}
11.96\\2.81\end{array}$ &104& 108.1&99.3&  4
\medskip\\
56731& H/S &7.07$\pm$0.68& -47.33$\pm$3.04& -26.48$\pm$0.5& & & 297.3$\pm$4.1& &
1.1$\pm$0.37& 6.07$\pm$0.08&71&&&\\
&F& 6.89$\begin{array}{l}7.67\\6.18\end{array}$&-48.53$\pm$3.9& -26.26$\pm$2.2&
7.91$\pm$1.3& 81.3$\pm$30.6& 298.1$\pm$19.8&
358.5$\pm$18.8&1.15$\begin{array}{l}1.49\\0.86\end{array}$&7.36$\begin{array}{l
}9.68\\4.31\end{array}$ &34& 170.6&25.9&  0\\
&N& 7.27$\begin{array}{l}8.16\\6.48\end{array}$&-44.94$\pm$3.7& -29.18$\pm$2.5&
8.97$\pm$1.6& 69.2$\pm$29.1& 332.2$\pm$21.2&
334.3$\pm$18.1&1.23$\begin{array}{l}1.62\\0.91\end{array}$&7.47$\begin{array}{l
}10.53\\3.79\end{array}$ &37& 146.3&37.9&  0\\
&A& 7.09$\begin{array}{l}7.80\\6.44\end{array}$&-47.69$\pm$3.2& -27.24$\pm$2.0&
7.88$\pm$1.2& 77.6$\pm$26.4& 308.8$\pm$17.6&
347.8$\pm$15.9&1.11$\begin{array}{l}1.41\\0.86\end{array}$&7.04$\begin{array}{l
}9.13\\4.34\end{array}$ &74& 240.7&77.1&  0\\
&O&&&& 7.92$\pm$1.0&73.7$\pm$24.3&301.0&351.3$\pm$12.6& 1.1$\pm$0.2\\
\hline
\end{tabular}
\end{sidewaystable*}

\addtocounter{table}{-1}
\begin{sidewaystable*}
\caption{(Continued.)}
\renewcommand{\arraystretch}{0.6}
\begin{tabular}{lccccccccccccccc}\hline
HIP&Sol&$\varpi$&$\mu_{\alpha}^*$&$\mu_{\delta}$&$a_0$&$i$&$\omega$&$\Omega$&$a
_0/\varpi$&$K_1$&N&$\chi^2$&$\chi^2$&$\alpha$\\
&&(mas)&(mas/y)&(mas/y)&(mas)&(\degr)&(\degr)&(\degr)&(AU)&(km/s)&&HIP&Orb&(\%)
\\\hline
\medskip\\
\multicolumn{4}{l}{\bf Strong Ba stars (cont.)}
\medskip\\
103546*& H/S &3.16$\pm$0.75& -2.89$\pm$0.82& -20.06$\pm$0.67& & & 47.9$\pm$2.6&
& 0.51$\pm$0.18& 7.77$\pm$0.11&70&&&\\
&F& 3.28$\begin{array}{l}4.20\\2.56\end{array}$&-3.08$\pm$0.9& -19.77$\pm$0.7&
1.44$\pm$0.9& 59.6$\pm$49.5& 141.4$\pm$52.9&
171.8$\pm$55.4&0.44$\begin{array}{l}0.91\\0.13\end{array}$&7.37$\begin{array}{l
}17.72\\0.45\end{array}$ &35& 32.4&28.8& 12\\
&N& 3.01$\begin{array}{l}4.17\\2.17\end{array}$&-2.63$\pm$1.1& -20.76$\pm$0.9&
1.99$\pm$1.0& 80.4$\pm$35.1& 103.4$\pm$31.8&
234.6$\pm$35.8&0.66$\begin{array}{l}1.37\\0.24\end{array}$&12.62$\begin{array}{
l}26.60\\3.29\end{array}$ &35& 27.1&21.8&  1\\
&A& 3.24$\begin{array}{l}4.13\\2.55\end{array}$&-3.01$\pm$0.8& -19.97$\pm$0.7&
1.42$\pm$0.8& 45.8$\pm$62.7& 132.9$\pm$82.6&
185.1$\pm$87.6&0.44$\begin{array}{l}0.87\\0.15\end{array}$&6.08$\begin{array}{l
}16.93\\0.00\end{array}$ &72& 54.1&49.8& 26
\medskip\\
110108& H/S &4.59$\pm$1.18& 6.02$\pm$2.01& 21.09$\pm$0.74& & & 73.65$\pm$14& &
0.74$\pm$0.26& 5.1$\pm$0.06&25&&&\\
&F& 6.05$\begin{array}{l}7.55\\4.84\end{array}$&13.29$\pm$6.9& 18.75$\pm$1.9&
7.16$\pm$4.8& 92.4$\pm$16.2& 151.1$\pm$29.4&
118.3$\pm$15.7&1.18$\begin{array}{l}2.48\\0.31\end{array}$&12.65$\begin{array}{
l}26.50\\3.12\end{array}$ &12& 13.3&4.6&  0\\
&N& 5.65$\begin{array}{l}7.50\\4.26\end{array}$&14.88$\pm$4.6& 18.71$\pm$1.4&
7.50$\pm$2.9& 79.5$\pm$15.3& 181.8$\pm$26.7&
125.0$\pm$18.5&1.33$\begin{array}{l}2.45\\0.61\end{array}$&13.96$\begin{array}{
l}26.25\\5.84\end{array}$ &13& 19.0&8.4&  1\\
&A& 5.34$\begin{array}{l}6.77\\4.21\end{array}$&10.16$\pm$4.0& 19.77$\pm$1.2&
5.30$\pm$2.6& 85.1$\pm$19.3& 172.3$\pm$28.4&
128.2$\pm$21.3&0.99$\begin{array}{l}1.87\\0.40\end{array}$&10.58$\begin{array}{
l}19.96\\3.95\end{array}$ &26& 27.1&18.9& 17\\
\hline
\end{tabular}

Remark: HIP 103546: Orbit not well defined
\end{sidewaystable*}

\addtocounter{table}{-1}
\begin{sidewaystable*}
\caption{(Continued.)}
\renewcommand{\arraystretch}{0.6}
\begin{tabular}{lccccccccccccccc}\hline
HIP&Sol&$\varpi$&$\mu_{\alpha}^*$&$\mu_{\delta}$&$a_0$&$i$&$\omega$&$\Omega$&$a
_0/\varpi$&$K_1$&N&$\chi^2$&$\chi^2$&$\alpha$\\
&&(mas)&(mas/y)&(mas/y)&(mas)&(\degr)&(\degr)&(\degr)&(AU)&(km/s)&&HIP&Orb&(\%)
\\\hline
\medskip\\
\multicolumn{4}{l}{\bf Tc-poor S stars}
\medskip\\
17296& H/S &6.27$\pm$0.63& -16.97$\pm$0.52& 19.34$\pm$0.61& & & 322$\pm$15& &
0.51$\pm$0.17& 8.47$\pm$0.2&74&&&\\
&F& 6.07$\begin{array}{l}6.87\\5.37\end{array}$&-16.11$\pm$0.6& 19.24$\pm$0.7&
1.72$\pm$0.7& 85.6$\pm$25.3& 335.7$\pm$26.4&
122.5$\pm$29.4&0.28$\begin{array}{l}0.44\\0.15\end{array}$&5.16$\begin{array}{l
}8.13\\2.42\end{array}$ &36& 32.9&22.7&  0\\
&N& 7.02$\begin{array}{l}7.85\\6.27\end{array}$&-18.14$\pm$0.6& 19.27$\pm$0.7&
3.77$\pm$0.9& 117.1$\pm$15.9& 346.5$\pm$17.4&
194.0$\pm$16.8&0.54$\begin{array}{l}0.74\\0.37\end{array}$&8.77$\begin{array}{l
}13.28\\4.97\end{array}$ &38& 71.7&35.2&  0\\
&A& 6.29$\begin{array}{l}6.98\\5.66\end{array}$&-16.98$\pm$0.5& 19.35$\pm$0.6&
2.30$\pm$0.6& 105.6$\pm$18.7& 334.3$\pm$19.0&
162.1$\pm$20.1&0.37$\begin{array}{l}0.52\\0.24\end{array}$&6.46$\begin{array}{l
}9.46\\3.63\end{array}$ &78& 103.3&83.9&  1\\
&O&&&& 1.80$\pm$0.6&115.6$\pm$22.7&322& 166.8$\pm$24.5& 0.28$\pm$0.1&\\
&T& 7.9$\begin{array}{l}8.2\\7.6\end{array}$&-17.9$\pm$0.21& 19.3$\pm$0.22&
4.50$\pm$0.29& 114.$\pm$4.3& 361.$\pm$4.5&
217$\pm$4.9&0.57$\begin{array}{l}0.63\\0.51\end{array}$&&154&&&
\medskip\\
25092& H/S &1.32$\pm$0.99& 48.18$\pm$0.79& -5.5$\pm$0.63& & & 232.51$\pm$33& &
0.53$\pm$0.18& 7.888$\pm$0.276&52&&&\\
&F& 1.73$\begin{array}{l}3.39\\0.88\end{array}$&48.49$\pm$1.0& -5.84$\pm$0.8&
2.26$\pm$1.0& 97.5$\pm$22.4& 355.7$\pm$36.3&
302.4$\pm$25.9&1.31$\begin{array}{l}3.69\\0.38\end{array}$&22.07$\begin{array}{
l}62.81\\5.54\end{array}$ &26& 19.9&16.4&  7\\
&N& 2.09$\begin{array}{l}3.84\\1.14\end{array}$&48.85$\pm$1.0& -5.85$\pm$0.9&
1.47$\pm$1.1& 106.7$\pm$41.1& 322.0$\pm$61.9&
299.9$\pm$48.1&0.71$\begin{array}{l}2.27\\0.10\end{array}$&11.50$\begin{array}{
l}38.68\\0.87\end{array}$ &26& 18.8&17.4& 48\\
&A& 1.96$\begin{array}{l}3.40\\1.13\end{array}$&48.58$\pm$0.9& -5.82$\pm$0.8&
1.84$\pm$0.9& 100.2$\pm$26.2& 351.4$\pm$40.7&
300.6$\pm$29.8&0.94$\begin{array}{l}2.46\\0.27\end{array}$&15.75$\begin{array}{
l}41.81\\3.67\end{array}$ &52& 36.8&34.1& 51\\
&T& 3.4$\begin{array}{l}4.3\\2.7\end{array}$&49.8$\pm$0.84&
-6.5$\pm$0.69&2$\pm$1.1& 115$\pm$21& 306$\pm$24&
297$\pm$20&0.7$\begin{array}{l}1.3\\0.3\end{array}$&&72&&&
\medskip\\
99312*& H/S &2.25$\pm$0.77& 7.43$\pm$0.54& 1.11$\pm$0.52& & & 128.63$\pm$1.1& &
0.37$\pm$0.13& 22.3$\pm$0.09&82&&&\\
&F& 3.78$\begin{array}{l}8.28\\1.72\end{array}$&7.44$\pm$0.6& 1.06$\pm$0.6&
2.02$\pm$1.9& 113.9$\pm$41.8& 80.3$\pm$60.0&
253.4$\pm$42.4&0.53$\begin{array}{l}2.28\\0.01\end{array}$&14.57$\begin{array}{
l}68.01\\0.17\end{array}$ &40& 27.2&25.8& 44\\
&N& 4.54$\begin{array}{l}9.29\\2.21\end{array}$&6.88$\pm$0.7& 1.23$\pm$0.7&
2.96$\pm$2.1& 132.4$\pm$31.5& 80.6$\pm$59.4&
252.0$\pm$42.1&0.65$\begin{array}{l}2.30\\0.09\end{array}$&14.39$\begin{array}{
l}67.18\\0.76\end{array}$ &42& 32.7&29.9& 16\\
&A& 4.18$\begin{array}{l}7.98\\2.18\end{array}$&7.28$\pm$0.6& 1.03$\pm$0.5&
2.42$\pm$1.8& 122.5$\pm$32.0& 82.4$\pm$49.6&
255.1$\pm$34.3&0.58$\begin{array}{l}1.94\\0.08\end{array}$&14.56$\begin{array}{
l}57.92\\0.97\end{array}$ &84& 60.9&59.7& 82\\
\hline
\end{tabular}

Remark:\\
HIP 99312: orbit not well defined
\end{sidewaystable*}

Table~\ref{Tab:orbits} lists the astrometric and orbital parameters for the
reliable orbits according to the criteria discussed above. The results from the
different processing modes are collected in Table~\ref{Tab:orbits}, according
to the symbol given in column~2: H/S refers to the parameters from the Hipparcos
catalogue and from the spectroscopic orbit (on that line $a_0/\varpi$ is the
semi-major
axis in A.U. estimated from Eq.~\ref{Eq:a0varpi}), F refers to the processing of
the
IAD from FAST only, N to the IAD from NDAC only, A from the processing of the
combined FAST/NDAC data set, O to the orbital parameters from the DMSA/O, and
T to the parameters resulting from the TD.

Most of the retained orbits are indeed characterized by $\chi^2/(N-9)$ values
of the order of unity,
as expected. The first-risk errors $\alpha$ are not always close to 0, but if
the derived value for the semi-major axis is in good agreement with its 
expected value, that agreement has been considered as sufficient for retaining
the orbit.  The only cases where the reverse situation occurs (small $\alpha$ 
but discrepant $a_0/\varpi$) are the dwarf barium star HIP~60299, the mild
barium star HIP~36042 and the CH star HIP~53763. Although the orbit of
the latter is not well defined, it has been kept in our final
list to illustrate the large uncertainty on $\varpi$ resulting from an 
orbital period close to 1~yr (Sect.~\ref{Sect:comparison}).

The F, N and A solutions for the retained orbits are also generally in good
agreement, the only exceptions being the dwarf Ba stars HIP~62409 and
HIP~116233, and the mild Ba star HIP~117607.    
However, the model parameters
of these systems are highly correlated, and the
different measurement errors in the different data sets thus drive the 
solution in different directions. 
This statement may be expressed in a quantitative way using the
concept of {\it efficiency} $\epsilon$ introduced by \cite*{Eichhorn-1989} and
\cite*{Pourbaix-1999:a}. 
It is defined as 
\begin{equation}
\label{Eq:efficiency}
\epsilon = \sqrt[p]{\frac{\prod\nolimits^p_{k=1}
    \lambda_k}{\prod\nolimits^p_{k=1} q_{kk}}},
\end{equation}
where $\lambda_k$ are the eigenvalues of the covariance matrix  
of the estimated parameters, $q_{kk}$ are its diagonal elements, 
and $p$ denotes the number of parameters in the model. If $\epsilon$
is close to unity, there is obviously little correlation between the
parameters. For the combined NDAC+FAST solution, it amounts to 0.32,
0.21 and 0.42  in the case of HIP~62409,
HIP~116233 and HIP~117607, 
respectively, thus translating some
degree  of correlation between the model parameters.

In Table~\ref{Tab:orbits}, the uncertainty on $a_0/\varpi$ has been computed 
by combining the upper and lower limits on $a_0$ and $\varpi$, 
thus neglecting any possible
correlation between these two quantities 
(which is generally small -- except for
the three systems listed above -- as derived from the efficiency being close to
unity). 

\begin{figure*}[htb]
\resizebox{0.32\hsize}{!}{\includegraphics{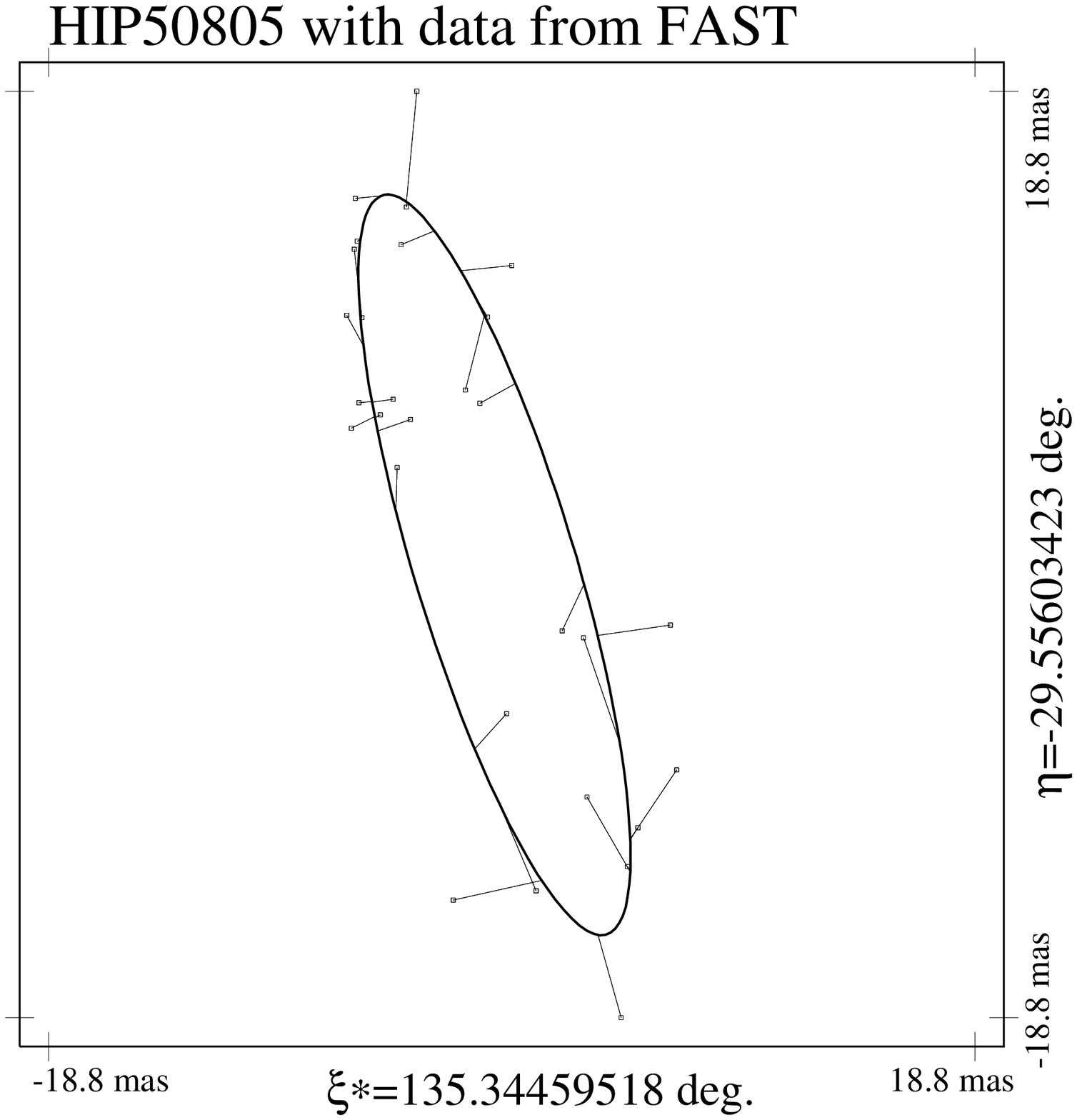}}\hfill
\resizebox{0.32\hsize}{!}{\includegraphics{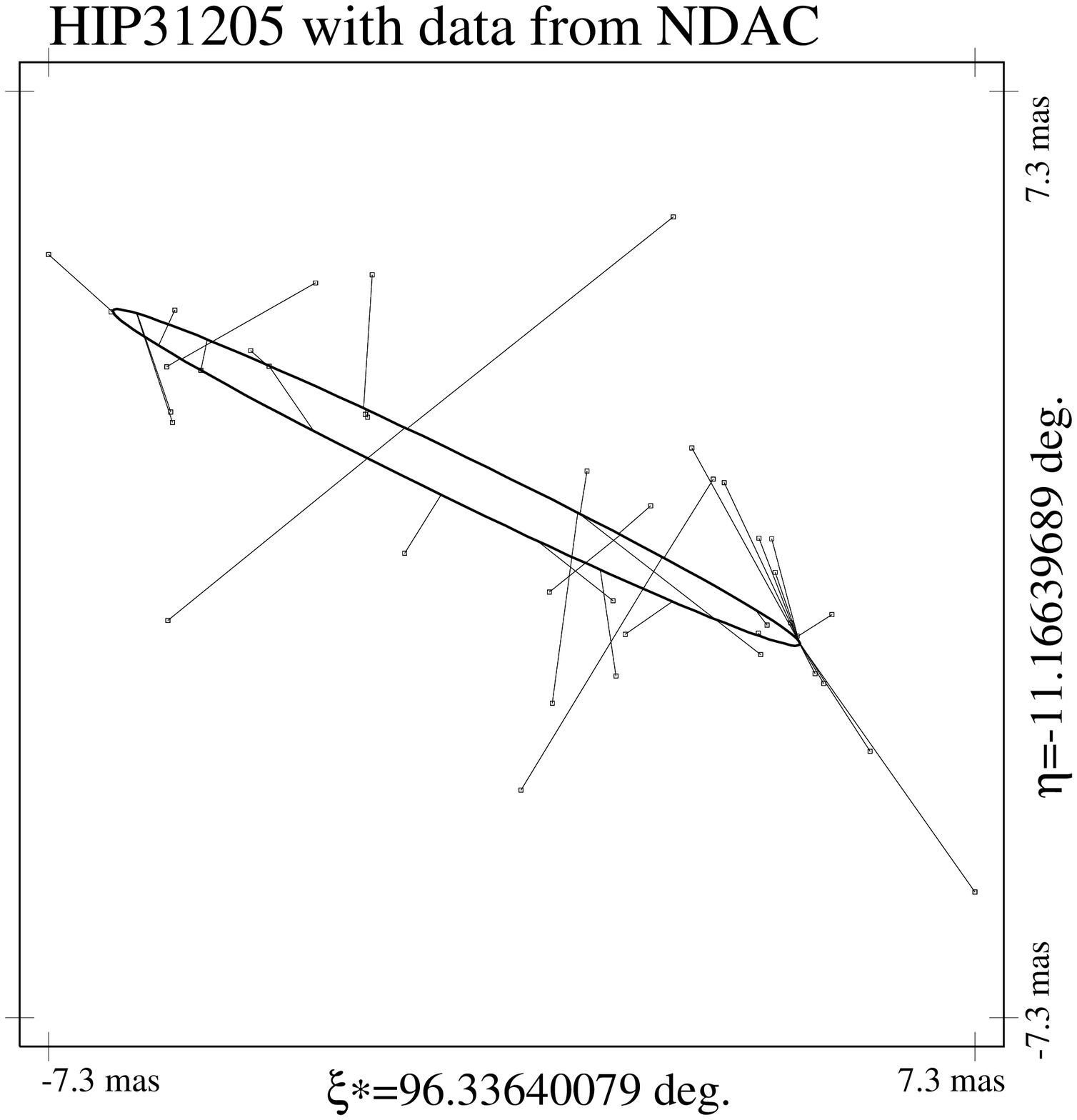}}\hfill
\resizebox{0.32\hsize}{!}{\includegraphics{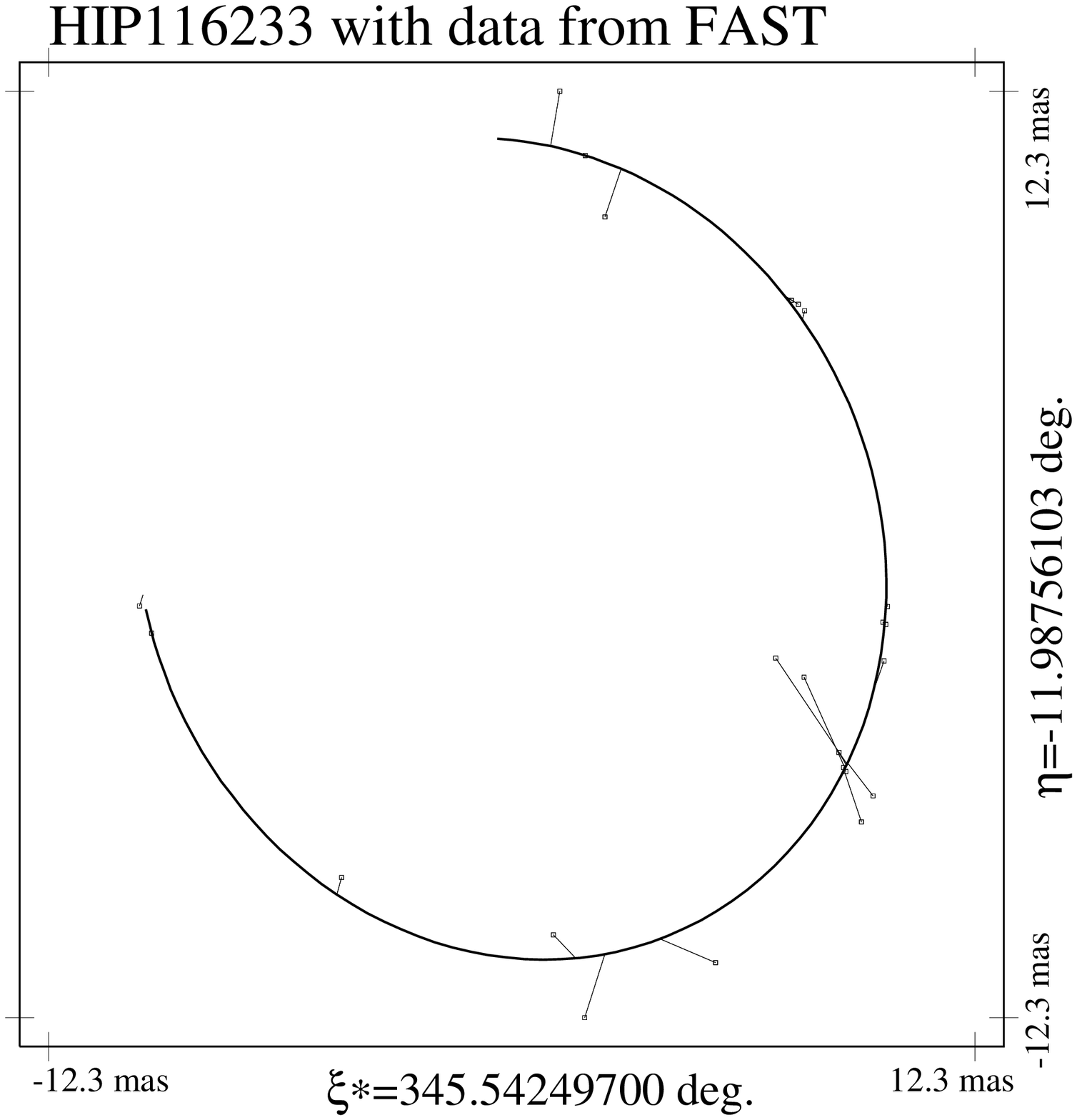}}
\par
\resizebox{0.32\hsize}{!}{\includegraphics{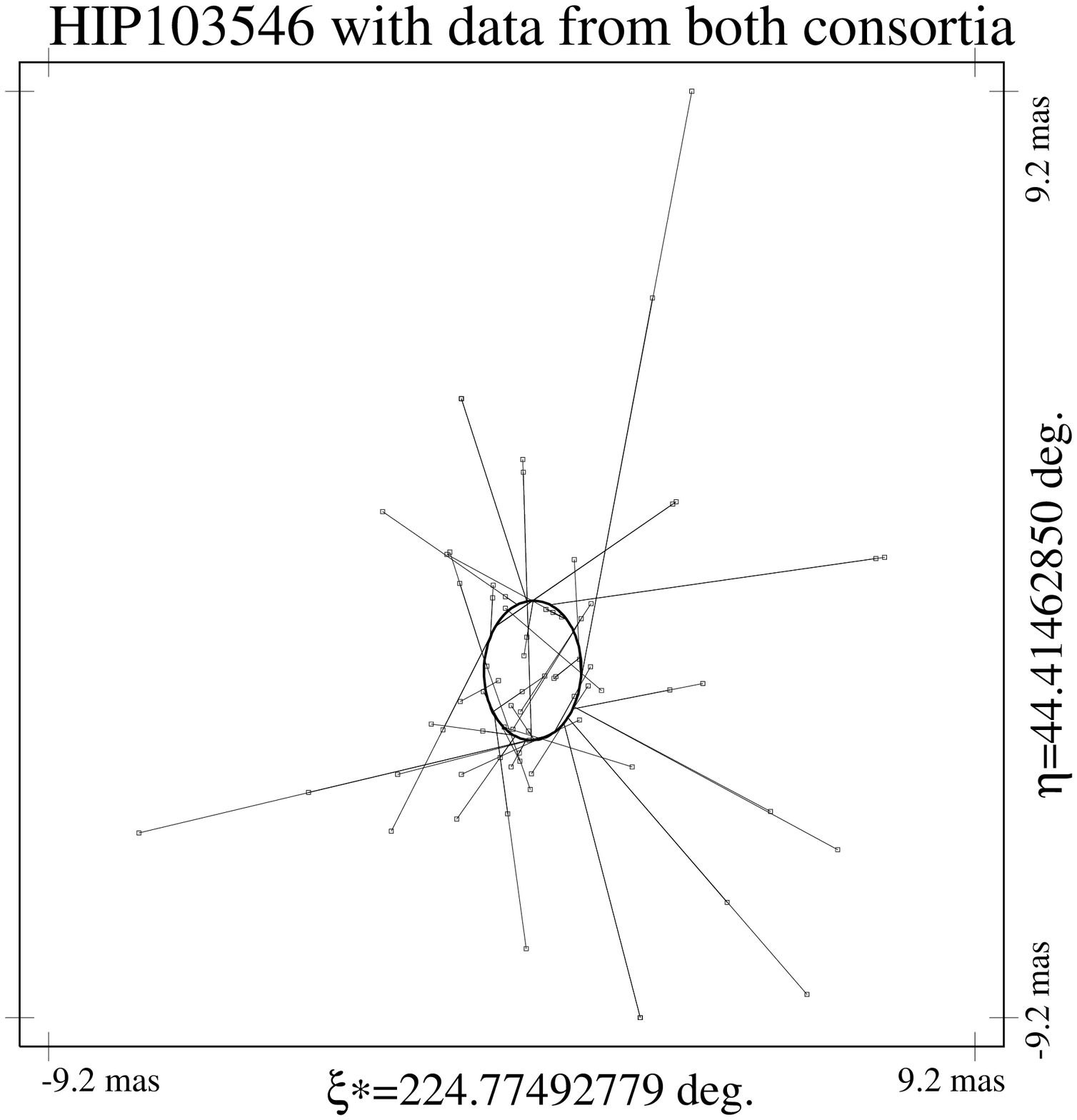}}\hfill
\resizebox{0.32\hsize}{!}{\includegraphics{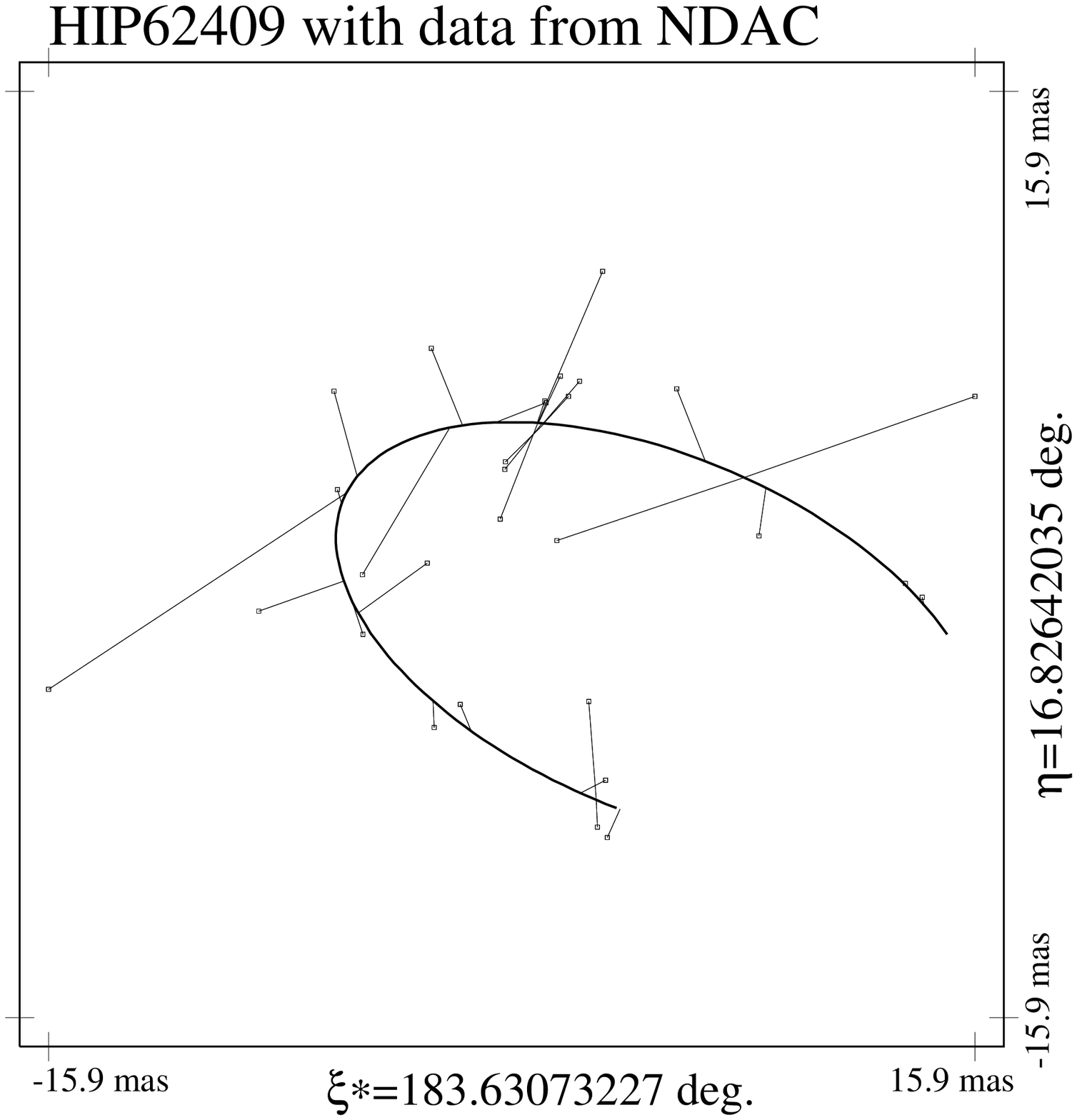}}\hfill
\resizebox{0.32\hsize}{!}{\includegraphics{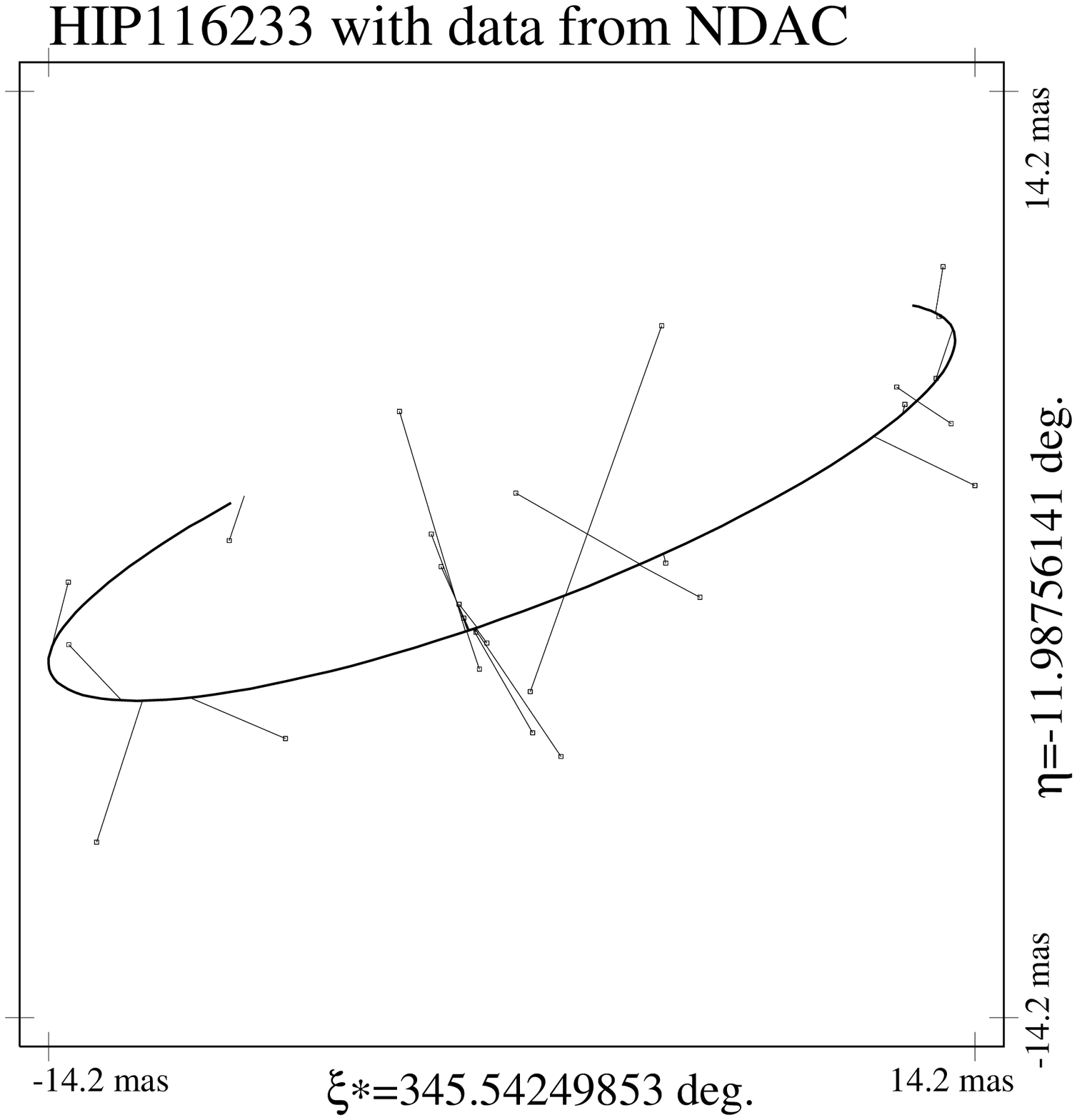}}
\par
\resizebox{0.32\hsize}{!}{\includegraphics{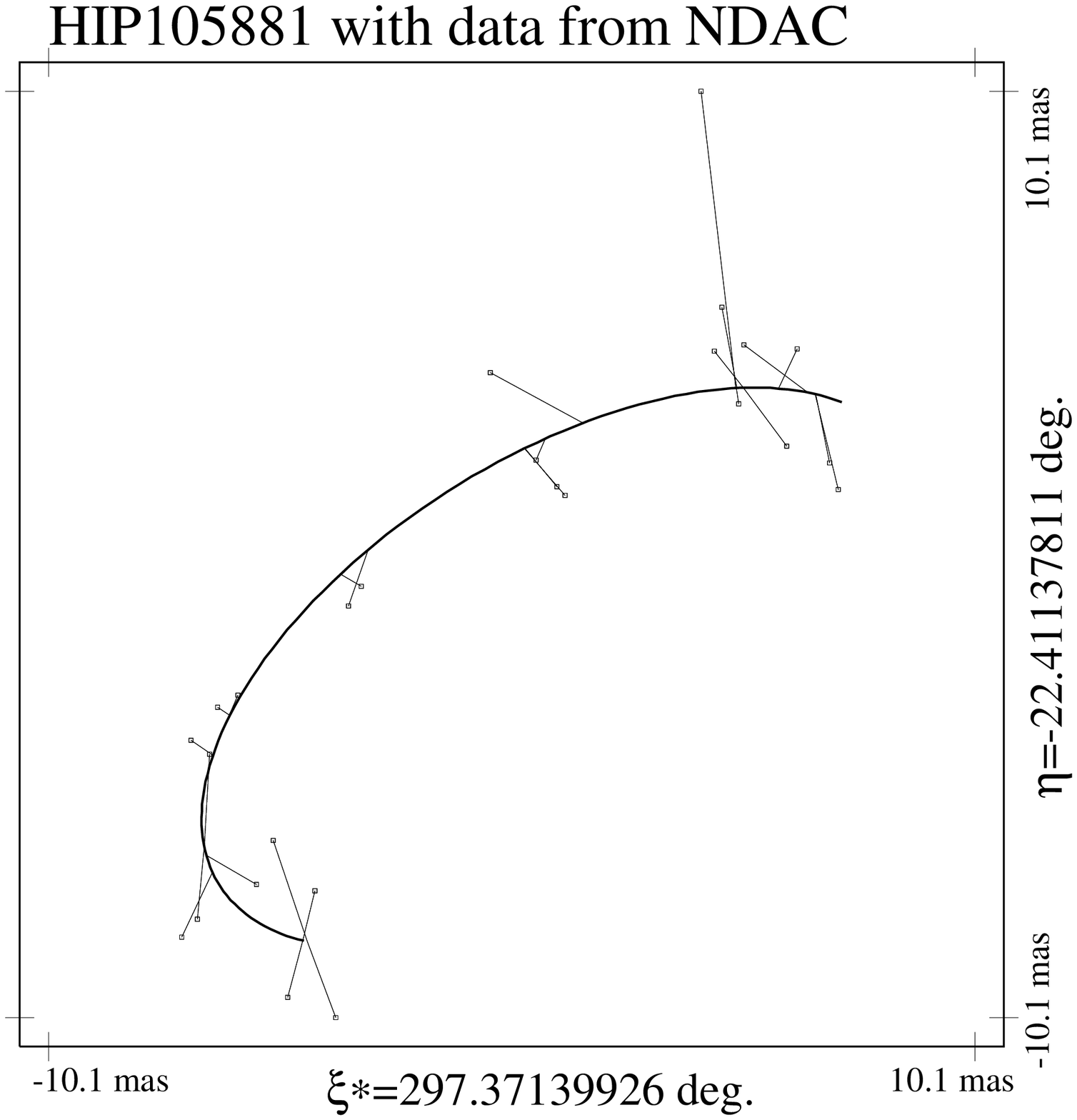}}\hfill
\resizebox{0.32\hsize}{!}{\includegraphics{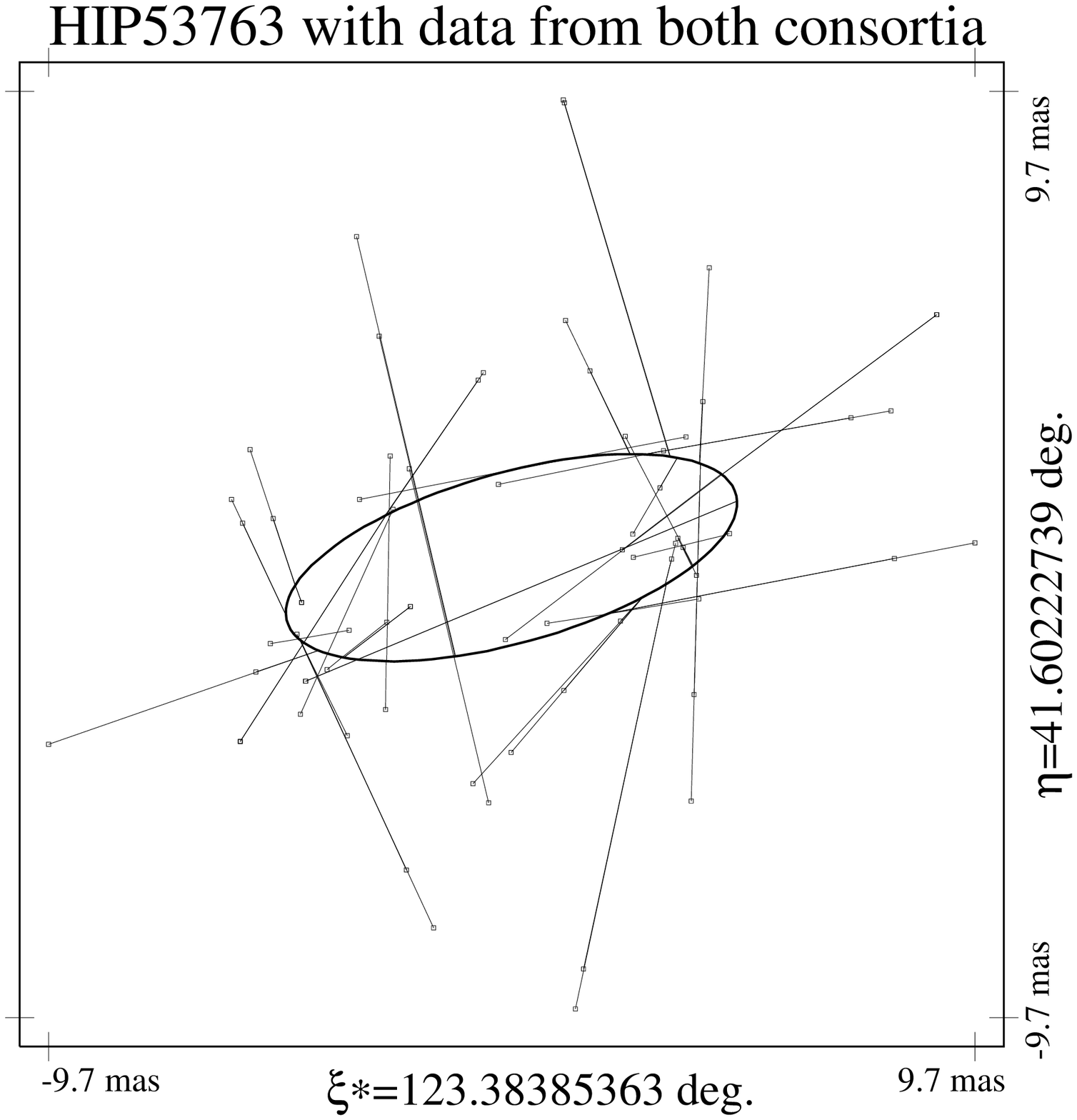}}\hfill
\resizebox{0.32\hsize}{!}{\includegraphics{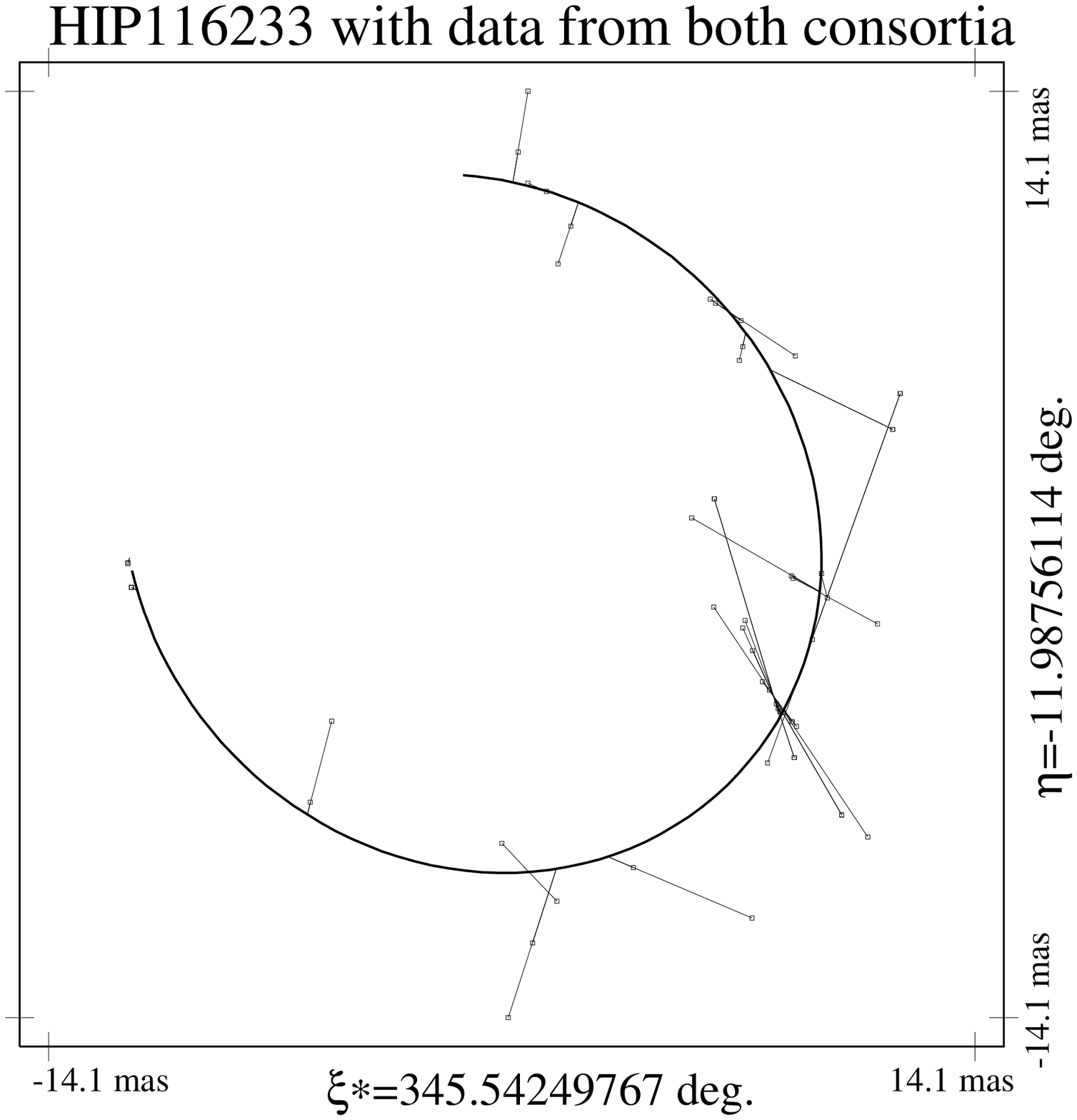}}
\caption{\label{Fig:arc}
The orbital arc on the plane tangent to the line of sight for some
representative orbits among those listed in
Table~\protect\ref{Tab:orbits}. The segments connect the computed
position on the orbit to the great circle (not represented,
perpendicular to the segment) corresponding to the
observed position (Hipparcos measurements are {\it one-dimensional}).
HIP 50805 is the only system for which the DMSA/O provides an orbital
solution from scratch; HIP~53763 has a small parallax (about 2~mas),
not very well determined since the orbital period is close to 1~yr; 
only the NDAC solution is acceptable for
HIP~62409; the orbit of HIP~103546 is at limit of what can be
extracted from the IAD; HIP~105881 is an example of an incomplete, albeit well
determined,  orbital arc; 
HIP~116233 is one case where the NDAC and FAST solutions are 
rather different
}
\end{figure*}

The orbital solutions derived in the present paper are too many to
display the astrometric orbit for all cases. A few representative cases among
the different subsets of Table~\ref{Tab:criteria} (orbital periods shorter
or longer than the duration of the Hipparcos mission, small or large
parallaxes\dots) have instead been 
selected and are presented in Fig.~\ref{Fig:arc}. 

\subsection{Comparison with DMSA/O solutions}

For HIP~17296 (Tc-poor S), 31205 (strong Ba) and 56731 (strong Ba), orbital
solutions are provided in the DMSA/O and were derived using spectroscopic
elements from the literature. HIP~50805 (dwarf Ba) is the only case in our
sample where an orbit could be derived from scratch by the Hipparcos consortia. 
For all these systems, the astrometric orbits derived by the methods
described in Sect.~\ref{Sect:numerical} are in excellent agreement with the
DMSA/O elements, thus providing an independent check of the validity of our
procedures. Further checks are presented in Sect.~\ref{Sect:checks}. 

The large number (23) of systems for which orbital solutions
could be extracted from the Hipparcos data (as compared to only 4 of those
already present in the
DMSA/O) illustrates the great potential that still resides in the  Hipparcos
IAD or TD.  

\subsection{Check of the astrometric orbit}
\label{Sect:checks}

Several checks are possible to evaluate the accuracy of the
astrometric elements derived in the present paper.
  
\begin{figure}[htb]
\resizebox{\hsize}{!}{\includegraphics{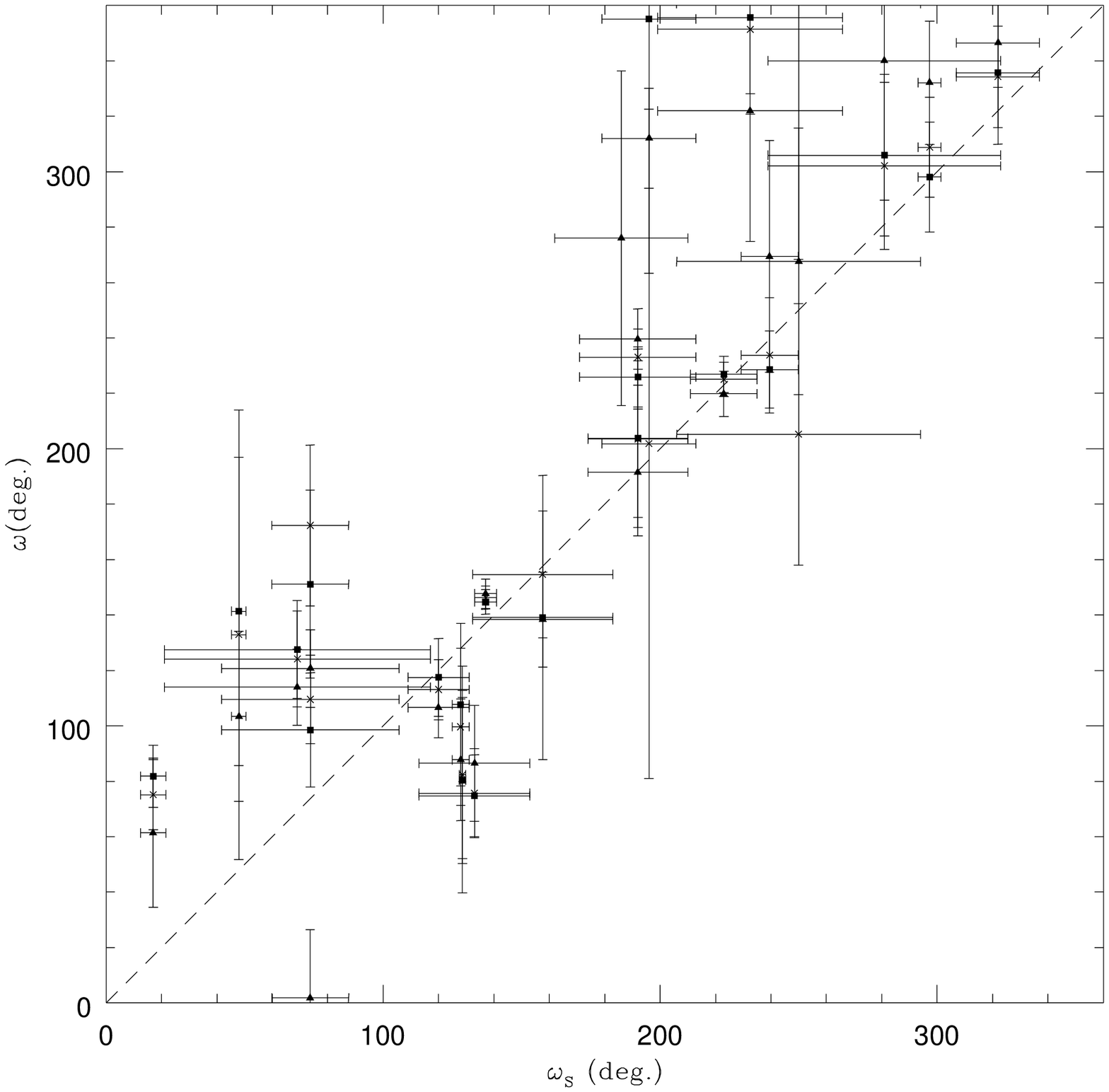}}
\caption{\label{Fig:omega}
Comparison of the astrometric (ordinate) and spectroscopic (abscissa)
determinations of the argument of periastron $\omega$ 
}
\end{figure}

First, it is possible to compare the astrometric and spectroscopic values 
of $\omega$, the argument of periastron.  In most cases, the two 
determinations agree within 2$\sigma$ (Fig.~\ref{Fig:omega}).  However,
even when the orbital period is shorter than the Hipparcos mission, the
$\omega$ derived from the IAD is seldom as precise as the spectroscopic one.

In a few cases, the spectroscopic orbit is assumed to be circular. In
that case,  the time $T$ of passage at periastron becomes meaningless, 
and is replaced by the time of the nodal 
passage (or, equivalently, the time of maximum radial velocity). This
is equivalent to setting   
$\omega$ equal to 0. Non-zero values for $\omega$ would correspond to
other conventions for the origin epoch.
Among our systems with acceptable orbits, two
have circular orbits: HIP~53763 (CH star) and HIP~52271 (strong barium star).
For these systems, our fit leads to values of $\omega$ significantly different
($2\sigma$) from 0, indicating that at the reference epoch, the star
is in fact far from the node where it was expected to be.

\begin{figure}[htb]
\resizebox{\hsize}{!}{\includegraphics{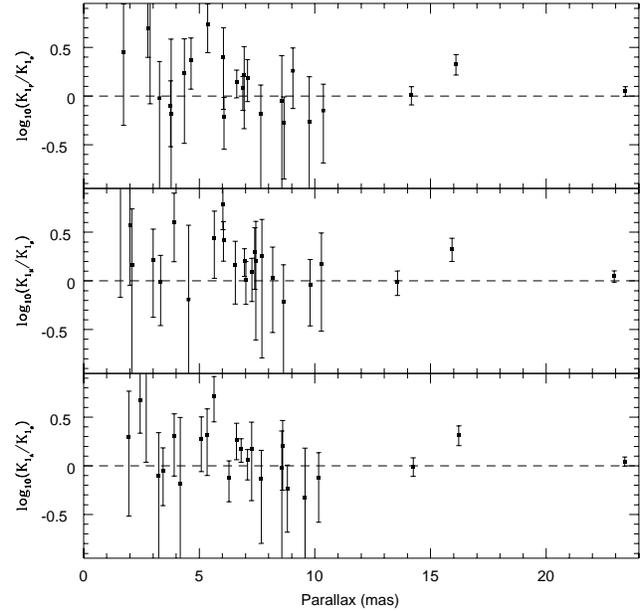}}
\caption{\label{Fig:K1}
Comparison of the semi-amplitude of the radial velocity curve derived from
the astrometry (using Eq.~\ref{Eq:check1}) and the spectroscopic value.
The three panels show the results derived from FAST, NDAC and FAST+NDAC (from 
top to bottom respectively)}
\end{figure}

\begin{figure}[htb]
\resizebox{\hsize}{!}{\includegraphics{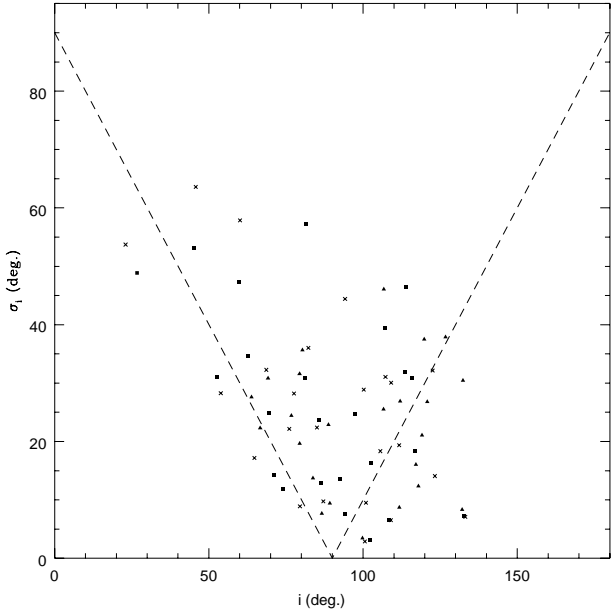}}
\caption{\label{Fig:sigi}
Uncertainty on the inclination $i$ as a function of $i$. All points
lying inside the region delineated by the oblique
dashed lines have orbital inclinations consistent with $i =
90$\degr. Filled triangles correspond to orbital solutions derived
from NDAC, filled squares to FAST and crosses to NDAC+FAST}
\end{figure}

\begin{figure}[htb]
\resizebox{\hsize}{!}{\includegraphics{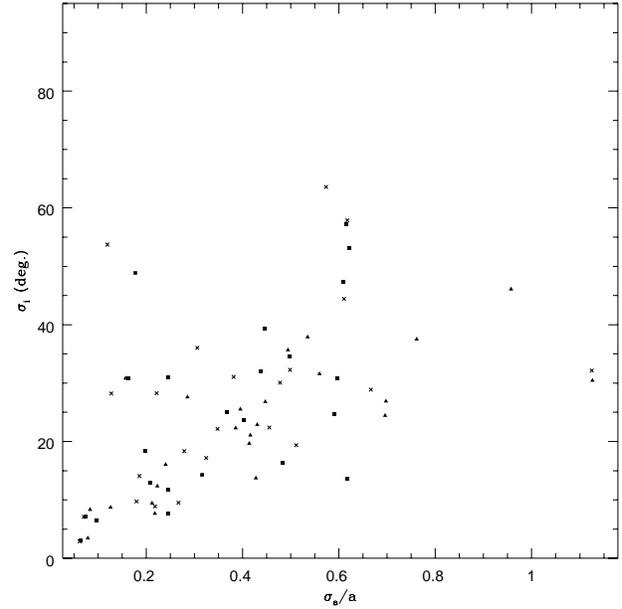}}
\caption{\label{Fig:sigasigi}
Correlation between the uncertainties on the inclination $i$ and
the semi-major axis $a_0$. The symbols have the same meaning as in
Fig.~\ref{Fig:sigi} 
}
\end{figure}

It is also possible to compare the astrometric value of $K_1$ (using in 
Eq.~\ref{Eq:check1} the value of $a$, $i$ and $\varpi$ from the astrometry 
and $e$ and $P$ from the spectroscopy) with the spectroscopic value.
Fig.~\ref{Fig:K1} shows that, even if $a_0/\varpi$ is well defined, the 
inclinations are generally not very accurately determined, thus leading to
uncertain values of $K_1$.  
This unfortunate property of $i$ is well illustrated
on Figs.~\ref{Fig:sigi} and \ref{Fig:sigasigi}.
One example where the accuracy of the astrometric value of $i$ must be 
questioned is the dwarf barium star HIP~105969: despite the fact that 
the astrometric
$a_0/\varpi$ ratio perfectly agrees with its estimate based on the
masses, the astrometric prediction of the semi-amplitude of the
radial-velocity variations differs by almost of factor of 2 as
compared to the actual spectroscopic value
(Table~\ref{Tab:orbits}). The only way to resolve that discrepancy is
to assume that the orbital inclination is largely in error.     

\begin{figure}
\resizebox{\hsize}{!}{\includegraphics{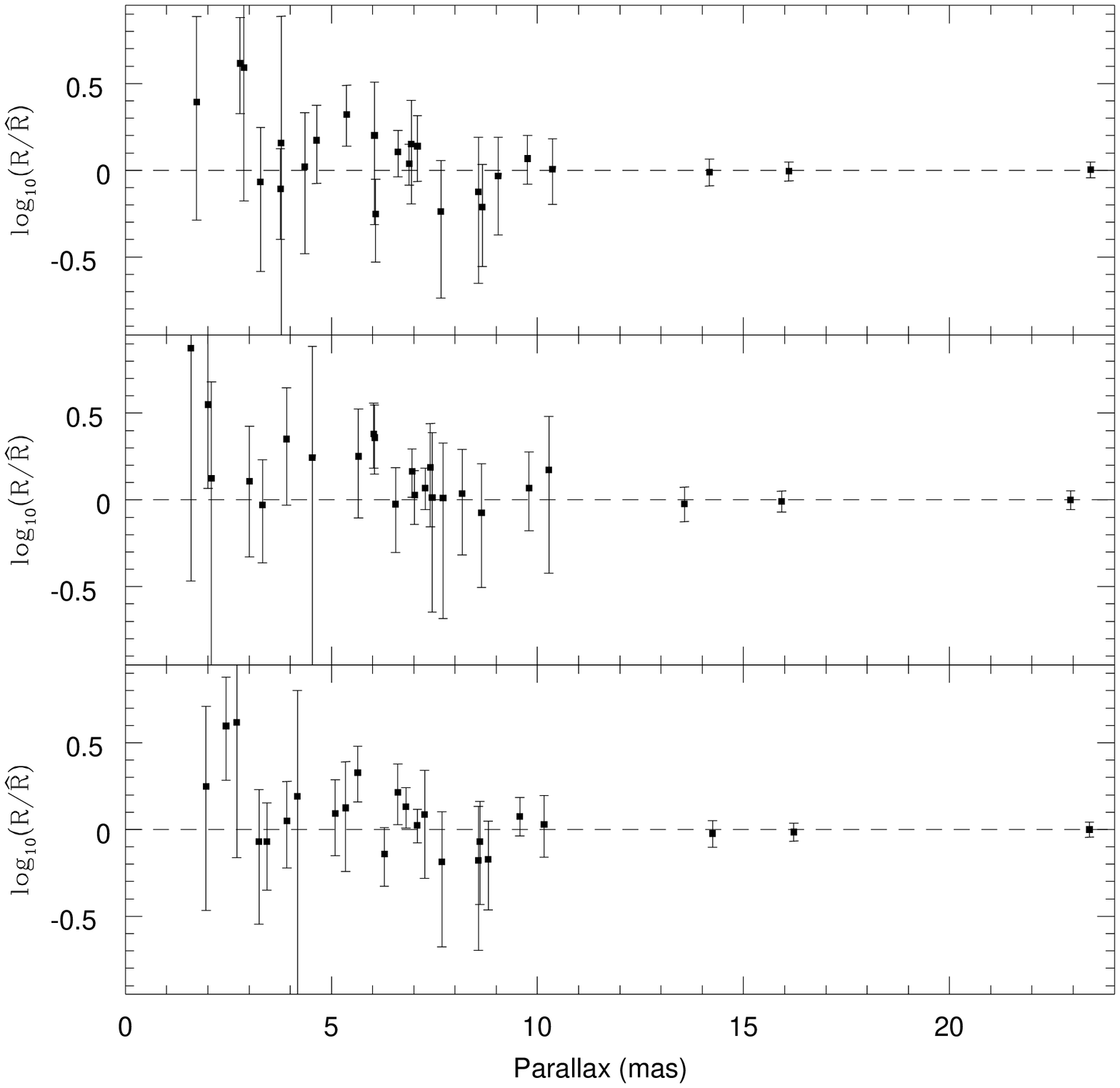}}
\caption{\label{Fig:a0varpi}
Comparison of the semi-major axis $R = a_0/\varpi$ derived from the astrometry
and from Eq.~\protect\ref{Eq:a0varpi}.  The upper panel displays the
results from the FAST data,  the middle panel the results from the NDAC data,
and the lower panel the results from  FAST and NDAC combined.  The error bars
assume that $M_1$, $M_2$ and $P$ are error-free}
\end{figure}

The semi-major axis $R = a_0/\varpi$ as derived from its astrophysical 
estimate (Eq.~\ref{Eq:a0varpi}) is compared to its astrometric value in 
Fig.~\ref{Fig:a0varpi}, and the two values are often consistent with each 
other.  Although the value of $a_0$ is likely to be affected by a positive
bias (i.e., a positive $a_0$ value is derived even when the data consist 
of pure noise, as clearly apparent from the astrometric $a_0/\varpi$ values
listed in Table~\ref{Tab:criteria}), this bias does not markedly affects 
the retained solutions displayed on Fig.~\ref{Fig:a0varpi}, except for 
solutions with $\varpi< 3$~mas, which all have $R/\hat{R} > 1$. Solutions 
for larger $\varpi$ values are almost equally distributed around unity.  

\subsection{Comparison TD vs IAD}
Among the 81 systems studied in this paper, 
22 have TD and only 5 belong to the list
of accepted orbits (HIP~17296, 25092, 31205, 5080, and 105881).  They hence 
can be used to check the consistency of the two orbit-determination methods.
The astrometric orbits derived from the TD are given in Table \ref{Tab:orbits}
as lines labeled with `T' in column 2.  
The agreement between the results obtained 
with the two  data types is quite good 
(within 1$\sigma$ error bars).  The number of 
TD measurements for those five systems is up to twice as large the
number of IAD measurements.  As a consequence, 
the confidence intervals of the parameters is systematically narrower.  One
should however keep in mind that more precise does not imply more accurate.
For instance, in the case of the mild Ba star HIP~105881, 
the 86 TD measurements yield $a_0/\varpi = 1.59$ 
(with 1$\sigma$ limits of 1.45 and 1.74)  as compared to $1.2\pm0.35$ 
estimated from Kepler's third law. The 54 IAD measurements from FAST and NDAC
combined yield a more accurate result of 0.99 (between 0.44 and 1.69). 

For the sake of completeness, we checked the astrophysical consistency of
the orbit derived for the 22 systems for which TD are available.  All systems
but the five already accepted were rejected.  In essence, this confirms that
the astrometric content of the TD is basically equivalent to the 
IAD and that nothing new can come out from TD
if IAD do not yield a reliable solution.

\section{Comparison of parallaxes and proper motions derived from single- or
double-star solutions}  
\label{Sect:comparison}

It has been stressed (\cite{Wielen-1997:b,Wielen-1999}) that the proper 
motions provided by the Hipparcos  catalogue may be systematically in error 
for long-period binaries (i.e. with $P > 3$~yr),  if those were not recognized
as such by the reduction consortia.  The orbital motion may then add to the 
actual proper motion, changing both its direction and modulus. The present 
sample  offers a good opportunity to evaluate the impact of this effect.   

\begin{figure}[htb] 
\resizebox{\hsize}{!}{\includegraphics{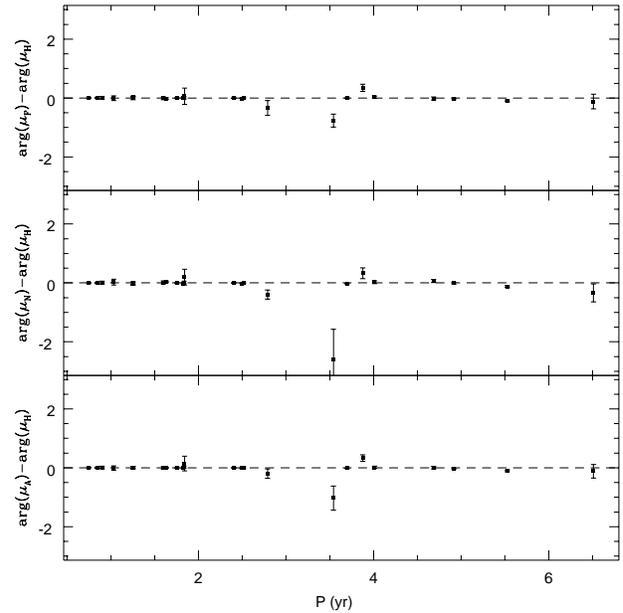}}
\caption{\label{Fig:argmu} 
Comparison of the position angle (expressed in radians) of the Hipparcos proper
motion  with that derived when account is made of the orbital motion (present
work), as a  function of the orbital period. The upper panel displays the
results from the FAST data,  the middle panel the results from the NDAC data,
and the lower panel the results from  FAST and NDAC combined. Error bars do not
include the uncertainty on the Hipparcos values} 
\end{figure}

\begin{figure}[htb] 
\resizebox{\hsize}{!}{\includegraphics{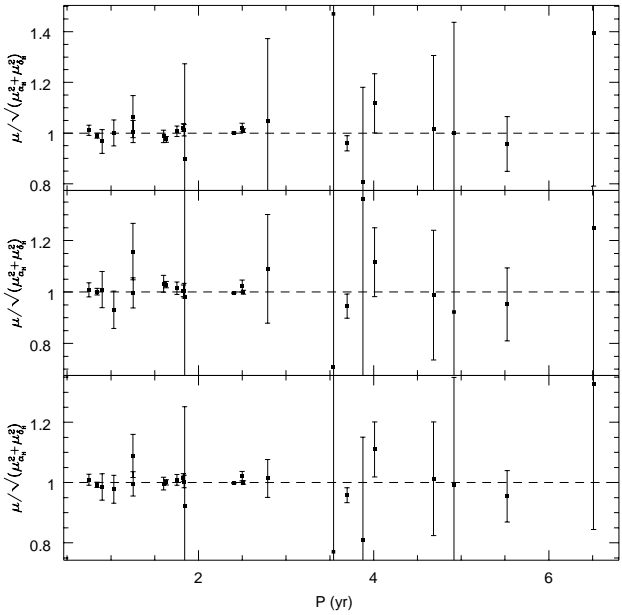}}
\caption{\label{Fig:mu} Same as Fig.~\protect\ref{Fig:argmu} for the 
proper-motion modulus} 
\end{figure}

Figs.~\ref{Fig:argmu} and \ref{Fig:mu} compare the position angle and 
modulus, 
respectively, of the Hipparcos proper motion with those derived when account is
made of  the orbital motion as in the present work. It is clearly apparent 
that the direction of the  proper motion listed in the Hipparcos catalogue is
correct, as expected, for orbits with  periods less than 3~yr, corresponding to
the duration of the Hipparcos mission. The  Hipparcos values are the less
accurate in the orbital-period range 3 to about 5~yr. The  position angle 
quoted by the Hipparcos catalogue may then be off by several dozens
degrees
(a good illustration of this situation is offered by the the strong Ba 
star HIP~110108 in Table~\ref{Tab:orbits}), 
though it differs generally by less than  $3\sigma$ from the correct value. 
For longer orbital periods, the orbital motion becomes negligible over the 
duration of the Hipparcos mission, and the proper motions are again rather 
well determined in the  Hipparcos catalogue.  The situation is almost 
identical for the the proper-motion modulus, except that the error bars 
now become very large for orbital periods longer than 3~yr. This situation 
translates the fact that the proper motion and the semi-major axis are 
strongly correlated (resulting in a large formal uncertainty on the proper 
motion), because the two motions are difficult to disentangle when the 
Hipparcos data sampled only a fraction of the orbit. This situation is 
encountered e.g. for HIP~104785 and 104732 (whose astrometric orbits
have not  been retained precisely because of the large uncertainty on
$a_0$ introduced by the strong correlation with $\mu$), 
and results in sometime large 
differences between the $a_0$ and $\mu$ values derived from the NDAC, FAST 
and NDAC+FAST data sets, since they are now very  sensitive to the measurement
errors.

\begin{figure}[htb] 
\resizebox{\hsize}{!}{\includegraphics{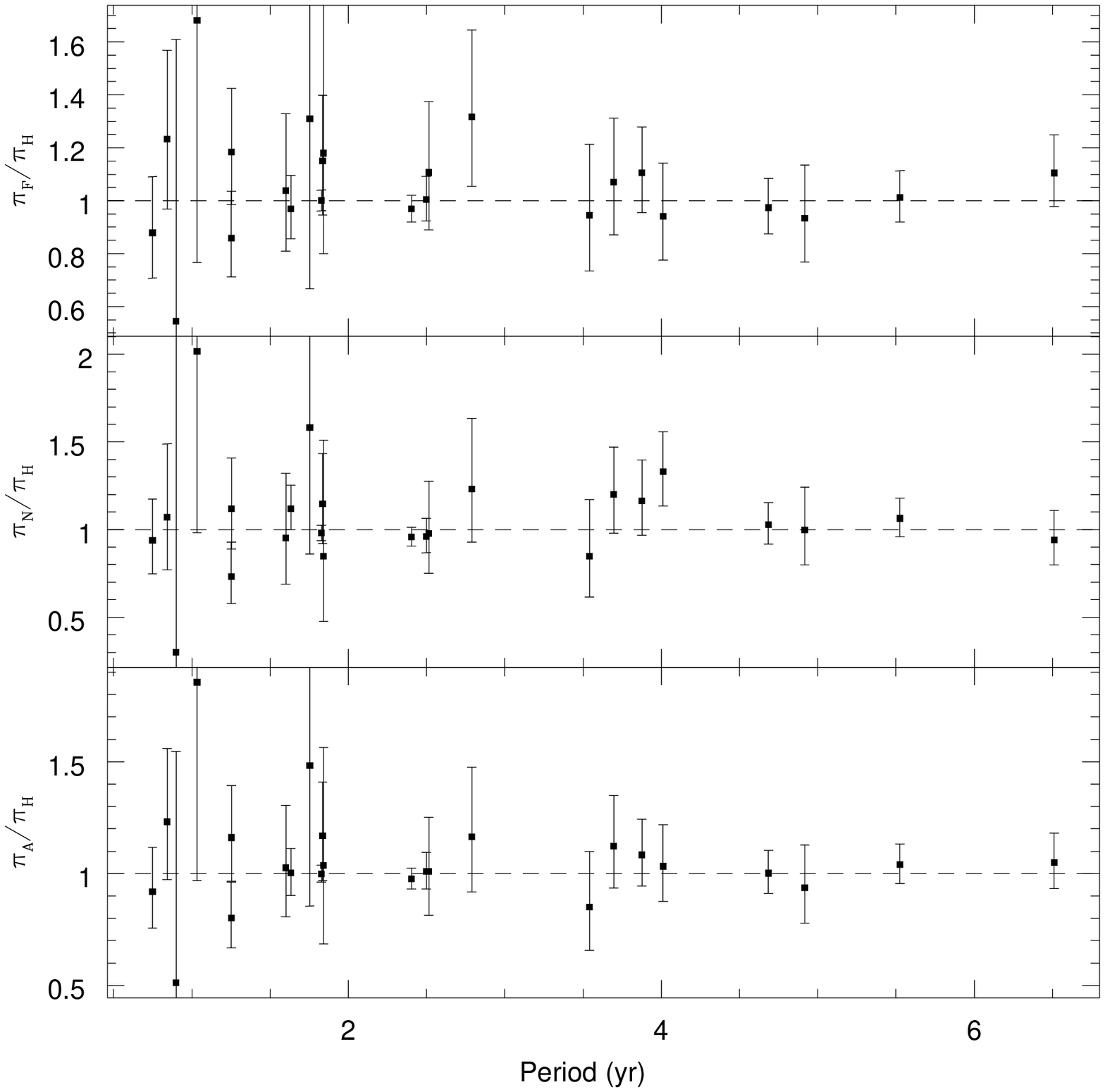}}
\caption{\label{Fig:piP} Comparison of the Hipparcos parallax with that
derived when account is made of the  orbital motion (present work), as a
function of the orbital period. As in  Fig.~\ref{Fig:argmu}, results
from FAST, NDAC and FAST+NDAC are presented in  the upper, middle and lower
panels, respectively. Error bars include only the uncertainty on 
the parallaxes derived from the
present work, and are computed according to Eq.~\ref{Eq:piprimerr}} 
\end{figure}

Fig.~\ref{Fig:piP} compares the parallax listed in the Hipparcos
catalogue with that derived in
the present  work, as a function of the orbital period. 
It turns out that the  agreement is good, except for systems with orbital
periods close to 1~yr. For those  systems, the parallax cannot be accurately
derived since the orbital motion and the  parallactic motion are strongly
entangled. This situation is encountered for HIP~17402, 53763 and
62827. 
Large error bars at periods different from 1~yr  correspond to systems
with parallaxes smaller than 3~mas.  

A few stars (HIP~8876, 29740, 29099, 32831, 36613) in our sample had negative
parallaxes in the Hipparcos catalogue.  The parallaxes we derive for these 
systems are slightly positive ($\varpi'$ was adopted to guarantee that
property), but the associated error bar encompasses zero, so
that these new parallaxes are just useless.  

\section{Masses}
\label{Sect:masses}

The test presented in Sect.~\ref{Sect:test} (Eq.~\ref{Eq:a0varpi}) 
to evaluate the astrophysical relevance of the orbital solution 
on the basis of $a_0/\varpi$ is 
in essence based on an a priori knowledge of the masses,
since Eq.~\ref{Eq:a0varpi} may in fact be rewritten as  
\[
Q =\frac{M_2^3}{(M_1 + M_2)^2} =\frac{(a_0/\varpi)^3}{P^2}.
\]

The orbital solutions that appear to be statistically significant on 
the basis of the F-test in almost all cases turned out to have a
semi-major axis 
$a_0/\varpi$ consistent with its expected value based on 
the estimated masses, the dwarf barium star HIP~60299 being however a
notable exception.  This agreement is well illustrated on 
Fig.~\ref{Fig:a0varpi}. 

Therefore, the present analysis does not add much to our previous knowledge 
of the masses, especially since the astrometric orbit only allows to eliminate 
$i$ from the mass function but does not give access to the individual masses. 

It was originally hoped that the present astrometric results might be
used to test the hypothesis which played 
a central role in the statistical analysis of the mass functions of
CPRS (\cite{Jorissen-1998}), 
namely that their $Q$ distributions are quite peaked, because they 
host a white dwarf companion.  
However, the error bar on $a_0/\varpi$ (and hence on $Q$) is too
large, even for the best determined barium-dwarf orbits
(Table~\ref{Tab:orbits}), to be able to draw a meaningful
$Q$-distribution from the astrometric orbits.

\section{Conclusions}

The availability of the Hipparcos Intermediate Astrometric Data and,
when they exist, the Transit Data, allow to improve upon the existing 
Hipparcos catalogue.  Once a new or revised spectroscopic binary orbit
becomes available,
it can be used to update the Hipparcos astrometric parameters and, sometime,
to derive the corresponding astrometric orbit.
In the latter case, the orbital inclination is obtained, thus allowing 
to improve our knowledge of the binary system.

The optimization method used in the present paper yields a solution in 
all cases, and it is therefore important to perform consistency
checks. One test which may be applied in all cases involves the  
comparison of the arguments of the periastron and of the
semi-amplitude of 
the radial-velocity curve, as derived from either the spectroscopic
orbit or the astrometric one. The comparison of the  
semi-amplitude of 
the radial-velocity curve involves however the inclination of the orbit which
is not always very well determined.

In this context, chemically-peculiar red stars (CPRS) like barium, Tc-poor S
and CH stars are interesting because
they provide their own consistency checks.  Indeed, beside the orbit
which can be determined as for any other spectroscopic binary, the
mass of both components may be guessed with some confidence. These
mass estimates based on astrophysical considerations may then be
compared to the mass of the system derived from the astrometric orbit
using Kepler's third law.

From a sample of 81 CPRS spectroscopic systems whose orbits became
available after completion of the Hipparcos catalogue, we have
derived 23 reliable astrometric orbits. This  
shows that the number of Hipparcos entries for which an orbital
solution may be obtained is much larger than
suggested by the number of existing entries in the DMSA/O.

Updated astrometric parameters from this particular sample have shown
that the Hipparcos-catalogue parallaxes are not reliable for systems with
orbital periods close to 1 yr. Similarly, the Hipparcos proper motions
are not reliable for unrecognized binaries with periods in the range 3
to about 8~yr, as already suspected by \cite*{Wielen-1999}.  

The orbit-determination methods based on the IAD and TD being
completely different, systems for
which the two sets of data are available can be used to assess the
consistency of the solutions derived from both sets with different
methods.  It turns out that the two sets of results are generally in
good agreement.

\begin{acknowledgements}
We thank P. North for communicating us the orbital elements of dwarf Ba 
stars in advance of publication.  We also thank two anonymous referees
for their very constructive comments.
\end{acknowledgements} 

\bibliographystyle{aabib}
\bibliography{articles,books}

\end{document}